\documentclass[12pt]{article}
\usepackage[top=1in,bottom=1in, left = 1in, right = 1in]{geometry}
\usepackage{amsmath,amssymb,amsfonts,amsthm,bbm}
\usepackage{epic,eepic,epsfig,longtable}
\usepackage{multirow,verbatim}
\usepackage{array}
\usepackage{graphicx}
\usepackage{floatrow}
\usepackage{subcaption}
\usepackage{paralist}
\usepackage{latexsym}
\usepackage{comment}
\usepackage{epsfig}
\usepackage{setspace}
\usepackage{CJK}
\usepackage{color}

\usepackage{geometry}
\usepackage{algorithm}
\usepackage{algpseudocode}
\usepackage[toc,page]{appendix}

\usepackage{rotating}
\usepackage[authoryear, round]{natbib}
\providecommand{\abs}[1]{\left\lvert#1\right\rvert}

\makeatletter
\def\singlespace{\def\baselinestretch{1}\@normalsize}

\numberwithin{equation}{section}

\renewcommand{\hat}{\widehat}

\renewcommand{\hat}{\widehat}
\newcommand{\bfm}[1]{\ensuremath{\mathbf{#1}}}

   \def\bA{\bfm A}  
   \def\bB{\bfm B}

\def\bff{\bfm f}    \def\FF{\mathbb{F}}
     
\def\bh{\bfm h}   \def\bH{\bfm H}  
   \def\bI{\bfm I}

   \def\bM{\bfm M}

\def\bq{\bfm q}   \def\bQ{\bfm Q}  
\def\br{\bfm r}   \def\bR{\bfm R}  \def\RR{\mathbb{R}}
   \def\bS{\bfm S}  
     
\def\bu{\bfm u}   \def\bU{\bfm U}  
   \def\bV{\bfm V}  
\def\bw{\bfm w}   \def\bW{\bfm W}  
   \def\bX{\bfm X}  
   \def\bY{\bfm Y}  
   \def\bZ{\bfm Z}

\def\calF{{\cal  F}}

\def\bzero{\bfm 0}

\newcommand{\bfsym}[1]{\ensuremath{\boldsymbol{#1}}}

\def\bbeta{\bfsym \beta}               
\def\bgamma{\bfsym \gamma}             \def\bGamma{\bfsym \Gamma}

\def\bmu{\bfsym {\mu}}                 
\def\bnu{\bfsym {\nu}}
\def\btheta{\bfsym {\theta}}           \def\bPsi  {\bfsym {\Theta}}
          
\def\bsigma{\bfsym \sigma}             \def\bSigma{\bfsym \Sigma}
        
\def\bomega {\bfsym {\omega}}          
\def\brho   {\bfsym {\rho}}

\def\bzeta{\bfsym {\zeta}}       \def\bvarrho{\bfsym {\varrho}}
\def\bPsi{\bfsym {\Psi}}
\def\bvartheta{\bfsym{\vartheta}}	
\def\bTheta{\bfsym{\Theta}}

\def\1{\bfsym{1}}	

              \def\hbtheta {\hat{\bfsym {\theta}}}

 \def\hlambda{\hat{\lambda}}

\DeclareMathOperator{\diag}{diag}
\DeclareMathOperator{\E}{E}

\def\newpage{\vfill\eject}

\def\today{\ifcase\month\or
	January\or February\or March\or April\or May\or June\or
	July\or August\or September\or October\or November\or December\fi
	\space\number\day, \number\year}

\newdimen\biblioindent    \biblioindent=30pt

 at 8truept

\newcommand{\beq}{\begin{equation}}
  \newcommand{\eeq}{\end{equation}}
\newcommand{\beqn}{\begin{eqnarray}}
  \newcommand{\eeqn}{\end{eqnarray}}
\newcommand{\beqnn}{\begin{eqnarray*}}
  \newcommand{\eeqnn}{\end{eqnarray*}}

\allowdisplaybreaks
\usepackage{verbatim}
\def\tilde{\widetilde}

\def\FF{\mathcal{F}}
\def\[{\left [}  \def\]{\right ]} \def\({\left (}  \def\){\right )}
 
\def\hat{\widehat}

\newtheorem{definition}{Definition}[section]
\newtheorem{assumption}{Assumption}[section]
\newtheorem{thm}{Theorem}[section]

\theoremstyle{remark}
\newtheorem{remark}{Remark}[section]

\def \diag {\mathrm{diag}}  
\def \det {\mathrm{det}} 
\def \Tr { \mathrm{Tr}} 
\def\vech {\mathrm{vech}}

\def \vec{\mathrm{vec}}

\usepackage{xcite}

\usepackage{xr}

\makeatletter

\newcommand*{\addFileDependency}[1]{

  \typeout{(#1)}

  \@addtofilelist{#1}

  \IfFileExists{#1}{}{\typeout{No file #1.}}

}

\makeatother

\newcommand*{\myexternaldocument}[1]{%

    \externaldocument{#1}%

    \addFileDependency{#1.tex}%

    \addFileDependency{#1.aux}%

}

\myexternaldocument{FOGI_appendix}

\graphicspath{{figures/}{./images}}

\title{Factor Overnight GARCH-It\^o Models}
\author{Donggyu Kim$^{1,\dagger}$, Minseog Oh$^1$, Xinyu Song$^2$, and Yazhen Wang$^3$  \\
$^1$College of Business, \\
Korea Advanced Institute of Science and Technology \\
$^2$School of Statistics and Management, \\
Shanghai University of Finance and Economics\\
$^3$Department of Statistics,\\
University of Wisconsin-Madison
}

\begin{document}
\maketitle

\begin{doublespace}
$\dagger$ Seoul, 02455, Republic of Korea, Email: donggyukim@kaist.ac.kr, Tel: +82-10-6399-9644
\clearpage
\begin{abstract}
This paper introduces a unified factor overnight GARCH-It\^o model for large volatility matrix estimation and prediction. 
To account for whole-day market dynamics, the proposed model has two different instantaneous factor volatility processes for the open-to-close and close-to-open periods, while each embeds the discrete-time multivariate GARCH model structure. 
To estimate latent factor volatility, we assume the low rank plus sparse structure and employ nonparametric estimation procedures.
Then, based on the connection between the discrete-time model structure and the continuous-time diffusion process, we propose a weighted least squares estimation procedure with the non-parametric factor volatility estimator and establish its asymptotic theorems.
\end{abstract}

\noindent \textbf{Keywords:}
Factor model,
high dimensionality,
POET,
quasi-maximum likelihood estimation,
realized volatility matrix estimator,
overnight risk

\noindent \textbf{JEL classification:}  C13, C32, C53, C55, C58



\section{Introduction}

Volatility estimation and prediction are vibrant research areas in financial econometrics and statistics.
There exist two major streams for studying volatility processes. 
The traditional stream adopts discrete-time parametric econometric models and employs low-frequency observations, such as daily, weekly, or monthly asset returns. 
The discrete-time models employ GARCH structures to explain the dynamic evolution of the volatility process and are easy to implement in financial applications. 
The well-known models include the ARCH \citep{engle1982autoregressive} and the GARCH models \citep{bollerslev1986generalized} for the univariate case, and the VEC-GARCH \citep{bollerslev1988capital}, the BEKK \citep{engle1995multivariate}, the CCC-GARCH  \citep{bollerslev1990modelling}, and the DCC-GARCH models \citep{engle2002dynamic} for the multivariate case. 
On the other hand, the rather modern stream is constructed based on the continuous-time diffusion process and employs high-frequency observations, such as transaction-by-transaction stock prices.
Given high-frequency data, nonparametric realized volatility estimators are constructed to estimate integrated volatilities. 
Examples of such estimators include the following: 
two-time scale realized volatility \citep{zhang2005tale}, 
multi-scale realized volatility \citep{zhang2006efficient, zhang2011estimating}, 
wavelet estimator \citep{fan2007multi}, 
kernel realized volatility \citep{barndorff2008designing, barndorff2011multivariate}, 
pre-averaging realized volatility \citep{christensen2010pre, jacod2009microstructure}, 
quasi-maximum likelihood estimator \citep{ait2010high, xiu2010quasi}, 
local method of moments \citep{bibinger2014estimating},
and robust pre-averaging realized volatility \citep{fan2018robust}.
For parametric models, 
\citet{nelson1990arch} and \citet{duan1997augmented} developed GARCH diffusions.
\citet{ghysels19965} explained It\^o process for multivariate stochastic volatility (MSV), and \citet{gourieroux2006continuous} developed a new multivariate diffusion process for covarince, called Wishart Autoregressive (WAR) process.
Recently, researchers have developed volatility dynamic models which employ nonparametric realized volatility estimators from high-frequency data to make inferences for parametric discrete-time models at the low-frequency \citep{asai2015forecasting, engle2006multiple, hansen2012realized, kim2016unified, shephard2010realising, song2021volatility}.
For a finite number of assets, their empirical studies showed that  combining the low- and high-frequency methods  helps improve the performance of volatility prediction. 
However, in financial applications, we often run into a large number of assets, which causes the curse of dimensionality. 
Thus, in this paper, we investigate an estimation procedure to unify the discrete-time and continuous-time models appropriately under the high-dimensional set-up for the purpose of large volatility matrix estimation and prediction.

To model high-dimensional and high-frequency data, we often employ approximate factor diffusion processes whose volatility matrix consists of the low-rank factor volatility and sparse idiosyncratic volatility matrices  \citep{ait2017using, fan2018robust, kim2019factor, kong2018systematic}. 
Under this set-up, to account for market volatility dynamics, \citet{kim2019factor} further described eigenvalue processes of the latent factor volatility matrix, based on a unified GARCH-It\^o model structure \citep{kim2016unified}. 
In their proposal, the daily integrated eigenvalues of the latent factor volatility matrix process have past squared factor returns as the innovations, which helps capture the common market dynamics.
Their proposed estimation procedure  provides more accurate results by harnessing  high-frequency financial data, and  the GARCH structure provides a method to predict future daily integrated volatility with its conditional expectation. 
However, these parametric models face the non-availability problem of high-frequency financial data during the market close-to-open period.
Thus, they are usually developed only based on the open-to-close period, which results in underestimated market risk.
Recently, to account for the whole-day market dynamics, for a single asset, \citet{kim2021overnight} introduced the overnight GARCH-It\^o model, which has two different instantaneous volatility processes for the open-to-close and close-to-open periods, respectively.
Empirical studies have supported incorporation of the close-to-open periods for explaining the whole-day market dynamics  \citep{kim2021overnight}. 
To analyze the risk contagion between the U.S. and China stock markets,  \citet{oh2023effect} proposed the overnight diffusion process, which can accommodate the fact that volatility transmission occurs through overnight volatility processes.
Unlike the case of a single asset, to account for whole-day market dynamics of large volatility matrices, we have to deal with two different types of data sources: low-frequency and high-frequency data, as well as the curse of dimensionality.

In this paper, we propose a large volatility matrix estimation and prediction procedure based on  high-dimensional factor It\^o diffusion processes. 
This approximate factor model decomposes the large volatility matrix into the latent factor volatility matrix that has low-rank, and the idiosyncratic volatility matrix that is sparse. 
Based on the approximate factor model, to model a latent factor volatility matrix, we  propose a factor overnight GARCH-It\^o (FOGI) model.
For example, the instantaneous factor volatility process of the FOGI model is obtained by some quadratic interpolation of the BEKK(1,1) model introduced by \citet{engle1995multivariate}.
Specifically, for the open-to-close period, the instantaneous factor volatility process has the current integrated factor volatility as an innovation, whereas for the close-to-open period, the instantaneous factor volatility process has the current squared log factor returns as an innovation.
Then, the conditional integrated factor volatility is expressed as a function of historical open-to-close integrated factor volatility matrices and squared close-to-open log factor returns.
To make inferences for the parameters of the FOGI model, we introduce a weighted least squares estimation procedure that assigns different weights to the open-to-close and close-to-open volatilities.
To do this under this high-dimensional set-up, we encounter an additional step of identifying and estimating the latent factor loading and factor volatility matrices. 
Given the proposed FOGI model structure, the factor loading matrix can be estimated consistently with the eigenmatrix or the group membership matrix \citep{ait2015using, kim2018large, kim2019factor, kim2022unified}. 
For example, we obtain the realized volatility matrix estimator based on open-to-close high-frequency data and use it as a proxy for the open-to-close conditional integrated volatility, whereas we use squared close-to-open returns as a proxy for the close-to-open conditional integrated volatility. 
These proxies have heterogeneous covariances that are related to the accuracy of the proxies. 
That is, we assume that the close-to-open and open-to-close volatilities have different dynamic structures.
To reflect this, we first estimate sample covariances for the close-to-open and open-to-close proxies, and then assign different weights based on the sample covariances to evaluate the proposed weighted squares loss function.
On the other hand, to account for the inherent sparse structure in the idiosyncratic volatility matrix, we employ the principal orthogonal component thresholding (POET) procedure introduced by \citet{fan2013large, fan2016incorporating} to estimate the idiosyncratic volatility matrix. 
By combining the estimators for the future factor and idiosyncratic volatility matrices, we introduce an estimator for the conditional expectation of the future large volatility matrix. 
The asymptotic theorems for the proposed estimator are established.

The rest of the paper is organized as follows.
Section \ref{SEC-MGARCH-Ito} introduces the FOGI model and studies its properties.
Section \ref{SEC-Parameter} develops the weighted least squares estimation procedure for model parameters and establishes its asymptotic theorems.
Section \ref{SEC-Large}  discusses how to harness the proposed model structure to estimate the future large volatility matrix,  and  we present the asymptotic properties of the proposed estimator.
In Section \ref{SEC-Simulation}, we conduct a simulation study to validate the finite sample performance of the proposed estimation methods. 
Section \ref{SEC-Empirical} applies the FOGI model to empirical data. 
The conclusion is presented in Section \ref{SEC-Conclusion}.
The  proofs are collected in the Appendix.


\section{Factor GARCH-It\^o models} 
\label{SEC-MGARCH-Ito}

First, let us fix the notations. 
The operator $\otimes$ denotes the Kronecker product, $\vec (\cdot)$ denotes the operator that stacks the columns of a matrix, and $\vech(\cdot)$  is a column vector obtained by vectorizing only the lower triangular part of a matrix. 
Moreover, $\vec^{-1} (\cdot)$ denotes the inverse operator of $\vec (\cdot)$.  
For any given $p_1$-by-$p_2$ matrix $\bM = \left(M_{ij}\right)_{i=1,\ldots, p_1, j=1,\ldots, p_2}$, denote its matrix spectral norm by $\|\bM\|_2$, its Frobenius norm by $\|\bM\|_F = \sqrt{ \Tr (\bM^{\top} \bM) }$, and $\| \bM \| _{\max} = \max_{i,j} | M_{ij}|$.
Let $C$ be a generic constant whose values are free of $\btheta$, $n, m,$ and $p$, and may change from occurrence to occurrence. 
We list the definitions of the notations in the Appendix.

 Let $\bX (t) = \(X_{1} (t), \ldots, X_{p}(t)\)^{\top}$ be the vector of true underlying log prices of $p$ assets at time $t$. 
To account for common market factors in the stock market, we employ the approximate factor models and assume the following factor It\^o diffusion process:
	\begin{equation*}
	d \bX(t) = \bU \bmu_t dt + \bU d \bff (t) + d \bu (t),
	\end{equation*}
where $\bmu_t \in \RR^{r}$ is a drift process, $\bU$ is a $p$-by-$r$ factor loading matrix and $r$ is the total number of market factors that is much smaller than $p$.
The drift has two different values for the open-to-close and close-to-open periods as follows:
\begin{eqnarray*}
	\bmu_t = \begin{cases}
  \bmu_H, & \text{ if } t \in  \( \lfloor t \rfloor,  \lfloor t \rfloor +\tau \], \\ 
\bmu_L,& \text{ if } t \in   \( \lfloor t \rfloor +\tau,  \lfloor t \rfloor +1 \] ,
\end{cases}
\end{eqnarray*}
where $\lfloor t \rfloor$ denotes the integer part of $t$ except that $\lfloor t \rfloor=t-1$ when $t$ is an integer, and $\tau$ is the length of the market open-to-close period.
Moreover, $\bff (t)$ and $\bu (t)$ are latent factor and idiosyncratic diffusion processes in the following: 
	\begin{equation*}
	d\bff (t) =  \bsigma_t^{\top} d \bB_t  \qquad \text{and} \qquad  d\bu(t) =  \bvartheta _t ^{\top} d \bW_t,
	\end{equation*}
where $\bsigma_t$ is an $r$-by-$r$ matrix, $\bvartheta_t$ is a $p$-by-$p$ matrix, and $\bB_t$ and $\bW_t$ are independent $r$-dimensional and $p$-dimensional Brownian motions, respectively. 
The stochastic process $\bsigma_t$ and $\bvartheta _t $ are defined on a filtered probability space $\( \Omega, \calF, \{\calF_t, t\in [0, \infty)\}, P \)$ with the filtration $\calF_t$ satisfying the usual conditions.
The instantaneous (or spot) volatility process of $\bff (t)$ is 
	\begin{equation*}
	\bSigma_t= \bsigma_t ^{\top} \bsigma_t.
	\end{equation*}
The daily integrated volatility matrix $\bGamma_n$ for the $n$th day consists of the factor volatility and idiosyncratic volatility matrices 
	\begin{equation*}
	\bGamma_{n} =   \bzeta_n + \bGamma_{n}^s,
	\end{equation*}
where 
	\begin{equation*}
	\bzeta_n = \bU \bPsi_n  \bU^{\top}, \quad \bPsi_n = \int_{n-1}^n \bsigma_t^{\top} \bsigma_t dt  \qquad \text{and} \qquad \bGamma_n^s = \int_{n-1} ^n \bvartheta_t  ^{\top} \bvartheta_t dt. 
	\end{equation*}
The latent factor process $\bff (t)$ corresponds to the systematic risk that is known to be undiversifiable, and it is important to model the rich dynamics in the factor volatility $\bPsi_n$ process. 
Thus, we impose a dynamic structure on the instantaneous factor volatility process to capture the market dynamics. 
On the other hand, the idiosyncratic process $\bu (t)$ corresponds to the firm-specific risk that can be usually mitigated through diversification. 
Thus, its corresponding volatility matrix $\bGamma^s_{n}= \(\Gamma_{n, ij} ^s \) _{i,j=1,\ldots, p}$ retains a sparse structure as follows:
 	\begin{equation} 
	\label{sparsity}
	\max_{1\leq j \leq p} \sum_{i=1}^p |\Gamma_{n, ij}^s| ^{\delta} |\Gamma_{n, ii}^s \Gamma_{n, jj}^s | ^{(1-\delta)/2} \leq M \pi(p) \quad \text{ a.s.},
	\end{equation}
where $\delta \in [0,1)$, $M$ is the positive bounded random variable, and the sparsity level $\pi(p)$ diverges very slowly, such as $\log p$ \citep{ait2015using, fan2019structured, fan2013large, fan2016incorporating, kim2018large, kim2022unified, kong2018systematic, shin2021adaptive}. 
We discuss its estimation in Section \ref{SEC-Idiosyncratic}.

When handling the whole-day market dynamics, we have two-types of data, such as the high-frequency data during the trading hours and the low-frequency data during the overnight period. 
Since low- and high-frequency data have different properties, we need to treat these two types of data differently.
To do this, we decompose the integrated factor volatility as follows:
\begin{equation*}
	\bPsi_n = \int_{n-1} ^n \bSigma_t dt=\int_{n-1} ^{\tau+ n-1} \bSigma_t dt +\int_{\tau+n-1} ^n \bSigma_t dt,
	\end{equation*}
where $\tau$ is the length of the open-to-close period.
Then, we impose two different dynamic structures on the open-to-close and close-to-open periods.
For example, during the open-to-close period, high-frequency data are observed so that volatility-related information can be reflected by the market price immediately, whereas during the close-to-open period, high-frequency data are not available and we usually cannot react to the overnight information immediately.
To represent this, we use the open-to-close integrated factor volatility and the close-to-open squared log factor return as innovations as follows.

\begin{definition}
\label{Def-MGARCH-Ito}
	We define the factor overnight GARCH-It\^o (FOGI) model.
	In the proposed model, the log factor price $\bff(t)$ and its instantaneous volatility process $\bSigma_t $ satisfy 
	\begin{eqnarray}	\label{def : UMGARCH}
		&& d \bff(t) =   \bsigma_t ^{\top}  d\bB_t, \cr
		&&\bSigma_t  = 
		\begin{cases}
			\bSigma_{\lfloor t \rfloor} + \( \frac{t-\lfloor t \rfloor}{\tau} \)^2   \bgamma_H \( \bomega _{H1} + \bSigma _{\lfloor t \rfloor} \)  \bgamma_H^{\top}\\
			 -  \frac{t-\lfloor t \rfloor}{\tau} \( \bomega _{H2} + \bSigma _{\lfloor t \rfloor}\) 
			+ \tau^{-1}\bbeta_H \int_{\lfloor t \rfloor} ^t \bSigma_s ds \bbeta_H^{\top}\\ 
			+ \tau^{-2} (\lfloor t \rfloor+\tau-t) \bnu \bZ_t  \bZ_t^{\top} \bnu ^{\top}, & \text{ if } t \in (\lfloor t \rfloor, \tau+\lfloor t \rfloor], \\ 
			\bSigma_{\tau+ \lfloor t \rfloor} + \(\frac{t-\lfloor t \rfloor-\tau}{1-\tau}\)^2 \bgamma_L \( \bomega _{L1}+  \bSigma _{\tau+ \lfloor t \rfloor} \)    \bgamma_L^{\top} \\
			-\frac{t-\lfloor t \rfloor-\tau}{1-\tau}  \( \bomega _{L2}+\bSigma _{\tau+ \lfloor t \rfloor} \)+ (1-\tau)^{-1}\bbeta_L \br_t \br_t^{\top}  \bbeta_L^{\top}, & \text{ if } t \in (\tau+ \lfloor t \rfloor , \lfloor t \rfloor +1] ,
		\end{cases}
	\end{eqnarray}
where 
$\bZ_t = \int_{\lfloor t \rfloor}^{t} d \tilde{\bB}_s$,
$\br_t= (r_{i,t})_{i=1,\ldots,r}=\int_{\lfloor t \rfloor+\tau}^{t} \bsigma_{s}^{\top} d\bB_s$,
$\tilde{\bB}_t$ is a $r$-dimensional standard Brownian motion with $d\tilde{\bB}_t d \bB_t =\brho dt$ a.s.,  
$\bomega_{H1}$, $\bomega_{H2}$, $\bomega_{L1}$, and $\bomega_{L2}$ are positive definite $r$-by-$r$ matrices, and  
$\bnu$, $\bbeta_H$, $\bbeta_L$, $\bgamma_H$ and $\bgamma_L$ are  $r$-by-$r$ symmetric matrices.
\end{definition}

The FOGI model has a continuous instantaneous volatility process concerning time $t$, and it obeys the standard BEKK (1,1) model structure with both the open-to-close integrated factor volatility matrix and the close-to-open squared log factor returns as innovations. 
For example, for $n \in \mathbb{N}$, the instantaneous volatility at the market opening is as follows: 
	\begin{eqnarray*} 
	\bSigma _n  &=& \bomega_L + \bgamma_L \bomega_H \bgamma_L ^{\top}  + \bgamma_L \bgamma_H  \bSigma _{n-1}  \bgamma _H^{\top}\bgamma_L^{\top} + \tau^{-1} \bgamma_L \bbeta_H \int_{n-1}^{\tau +n-1} \bSigma_t dt \bbeta_H^{\top} \bgamma_L^{\top}   \cr
	&& +(1-\tau)^{-1} \bbeta_L \br _{n} \br_{n}^{\top} \bbeta_L^{\top},
	\end{eqnarray*}
where $\bomega_H = \bgamma_H \bomega_{H1} \bgamma_H - \bomega_{H2}$, $\bomega_L = \bgamma_L \bomega_{L1} \bgamma_L - \bomega_{L2}$,
and, at the market closing, is as follows:  
	\begin{eqnarray*} 
	\bSigma _{\tau+n}  &=& \bomega_H + \bgamma_H \omega_L \bgamma_H ^{\top}  + \bgamma_H \bgamma_L  \bSigma _{\tau+ n-1}  \bgamma _L^{\top}\bgamma_H^{\top} + \tau^{-1}  \bbeta_H \int_{n}^{\tau+n} \bSigma_t dt \bbeta_H^{\top}    \cr
	&&        +(1-\tau)^{-1}\bgamma_H \bbeta_L \br _{n} \br_{n}^{\top} \bbeta_L^{\top} \bgamma_H^{\top}.
	\end{eqnarray*}
Moreover, the instantaneous volatility process can reflect the intraday U-shape trading pattern  \citep{admati1988theory, andersen1997intraday, andersen2018time, hong2000trading} by employing the quadratic interpolation form. 
Thus, the FOGI model has some quadratic interpolation term. 
This interpolation gives more weight to the current information dynamics (related with the squared term) and reduces the weight of the past information dynamics (related with the linear term). 
That is, at the open or close, the instantaneous volatility process has the BEKK form with the current available data, while between the open and close, it has their some weighted average.
The terms $\bgamma_H \( \bomega _{H1} + \bSigma _{\lfloor t \rfloor} \)  \bgamma_H^{\top}$ and $\bgamma_L \( \bomega _{L1}+  \bSigma _{\tau+ \lfloor t \rfloor} \)    \bgamma_L^{\top}$
help obtain the MA-type property.
The terms $\bbeta_H \int_{\lfloor t \rfloor} ^t \bSigma_s ds \bbeta_H^{\top}$ and  $\bbeta_L \br_t \br_t^{\top}  \bbeta_L^{\top}$ represent open-to-close and close-to-open innovations, respectively. 
The squared random noise term $\bZ_t \bZ_t^{\top}$ represents the random fluctuation, and we can reflect the current market volatility via the integrated factor volatility $\int_{\lfloor t \rfloor} ^t \bSigma_s ds$  and squared overnight factor returns $\br_t \br_t^{\top}$. 
We note that when we consider the single factor, that is, $r=1$, the FOGI model returns to  the overnight GARCH-It\^o model structure \citep{kim2021overnight}.
That is, the proposed FOGI model can be considered as the generalized version of the overnight GARCH-It\^o model.

For statistical inference, we study the integrated factor volatilities obtained from the FOGI model over consecutive integers, and during market opening and closing periods. 

\begin{thm} 
\label{Prop-IV}
We have the following integrated volatility structures for the proposed FOGI model. 
	\begin{itemize}
		\item[(a)]
		For $\det(\bbeta_H) \neq 0$ and $\| \bbeta _H\|_2<1$,  we have
			\begin{eqnarray} 
			&& \int_{n-1}^{\tau+n-1} \vec \( \bSigma_t  \) dt = \tau \bh_n^H (\btheta) + \bQ_{n}^H  \text{ a.s.},  \label{eq-Hv}\\
			&&\bh_n^H(\btheta)  =  \vec \( \bomega_{H}^g\) + \bR_H^g  \bh_{n-1} ^H (\btheta)  +  \frac{  \bA_H^g} {\tau}\int_{n-2}^{\tau +n-2} \vec \( \bSigma_t \) dt  + \frac{ \bB_H^g}{1-\tau}  \vec \( \br _{n-1} \br_{n-1}^{\top} \), \nonumber  
			\end{eqnarray}
		where 		
		$\bR_H= \bgamma_H \otimes \bgamma_H$, $\bB_H= \bbeta_H \otimes \bbeta_H$, $\bI_{r^2}$ is a $r^2$-dimensional identity matrix,
		$e^{\bB} = \sum_{k=0}^{\infty} \bB^{k}/ k!$,
		$\bvarrho_{H1}=\bB_H^{-1} (e^{\bB_H} -\bI_{r^2}) $,
		$\bvarrho_{H2}=\bB_H^{-2} (e^{\bB_H} -\bI_{r^2}-\bB_H ) $, $\bvarrho_{H3} =\bB_H^{-3} (e^{\bB_H} -\bI_{r^2}-\bB_H-\bB_H^2/2)$, 
		$\bvarrho_H=2\bvarrho_{H3}    \bR_H   +    \bvarrho_{H1} - \bvarrho_{H2} $,
		$\vec (\bomega_H^g) =  \(\bI_{r^2} - \bvarrho_H \bR_L \bR_H \bvarrho_H ^{-1} \)    \{ 2  \bvarrho_{H3} \bR_H  \\ \vec(\bomega_{H1}) -  \bvarrho_{H2}   \vec(\bomega_{H2})   + \(\bvarrho _{H2} - 2 \bvarrho_{H3}  \) \vec ( \bnu \bnu ^{\top}) \}    +  \bvarrho_H  \vec \(  \bomega_L \) + \bvarrho_H \bR_L \vec \(\bomega_H \) $,
		$\bR_H^g = \bvarrho_H  \bR_L \bR_H \bvarrho_H ^{-1}$,
		$\bA_H^g= \bvarrho_H  \bR_L \bB_H $,
		$\bB _H^g = \bvarrho_H  \bB_L$,
		$A_{k,ij} =\int _{n-1} ^{\tau +n-1} \frac{ (\tau+n-1-t)^{k+2} }{(k+2)\times k!} Z_{i,t}  dZ_{j,t} $,
		and 
		$$
		\bQ_n ^H = \frac{ \bnu \otimes \bnu }{ \tau^{ 2}}  \sum_{k=0}^{\infty} \tau^{-k} \bB_H ^k\vec \(  \(A_{k, ij} + A_{k, ji} \)_{i,j=1,\ldots, r}     \)
		$$ 
		is a martingale difference.
			
		\item[(b)]
		For $\det(\bbeta_L) \neq 0$ and $\| \bbeta_L \|_2<1$,  we have
			\begin{eqnarray*}
			&& \int_{\tau+ n-1}^{n} \vec \( \bSigma_t  \) dt = (1- \tau) \bh_n^L (\btheta) + \bQ_{n}^L  \text{ a.s.}, \cr
			&&\bh_n^L (\btheta)  = \vec \( \bomega_{L}^g\) + \bR_L^g  \bh_{n-1} ^L (\btheta)  +  \frac{  \bA_L^g} {\tau}\int_{n-1}^{\tau +n-1} \vec \( \bSigma_t \) dt  + \frac{ \bB_L^g}{1-\tau}  \vec \( \br _{n-1} \br_{n-1}^{\top} \),  
			\end{eqnarray*}
		where  $\bR_L= \bgamma_L \otimes \bgamma_L$, $\bB_L= \bbeta_L \otimes \bbeta_L$,
		$\bvarrho_{L1} =\bB_L ^{-1} (e^{\bB_L} -\bI_{r^2} )$, $\bvarrho_{L2}= \bB_L ^{-2} (e^{\bB_L} -\bI_{r^2}-\bB_L )$, $\bvarrho_{L3}= \bB_L ^{-3} (e^{\bB_L} -\bI_{r^2}-\bB_L - \bB_L ^2/2)$,
		$\bvarrho_L=2\bvarrho_{L3}    \bR_L   +    \bvarrho_{L1} - \bvarrho_{L2} $,		
		$\vec (\bomega_L^g) =  \(\bI_{r^2} - \bvarrho_L \bR_H \bR_L \bvarrho_L ^{-1} \)    \{ 2  \bvarrho_{L3} \bR_L   \vec(\bomega_{L1}) -  \bvarrho_{L2}   \vec(\bomega_{L2})  \}    +  \bvarrho_L  \vec \(  \bomega_H \) + \bvarrho_L \bR_H \vec \(\bomega_L \) $,
		$\bR_L^g = \bvarrho_L  \bR_H \bR_L \bvarrho_L ^{-1}$,
		$\bA_L^g= \bvarrho_L   \bB_H $,
		$\bB _L^g = \bvarrho_L \bR_H \bB_L$,
		$$
		\bQ_n^L = \sum_{k=1}^{\infty}   \( \frac{\bB_L} {1-\tau } \)^{k} \vec \(  \( \int _{\tau+ n-1} ^{n} \frac{ (n-t)^{k}}{ k !} r_{i,t}  d r_{j,t} +\int _{\tau+ n-1} ^{n} \frac{ (n-t)^{k}}{ k !} r_{j,t}  d r_{i,t}  \)_{i,j=1,\ldots, r} \)
		$$ 
		is a martingale difference.

		\item [(c)] For $\det ( \bbeta _H ) \neq 0$, $\det ( \bbeta _L ) \neq 0$, $\| \bbeta_H \|_2  <1$, and $\| \bbeta_L \|_2  <1$, we have
			\begin{eqnarray*}
			&& \int_{n-1}^{n} \vec \( \bSigma_t  \) dt = \bh_n (\btheta) + \bQ_{n}  \text{ a.s.}, \cr
			&&\bh_n (\btheta)  = \vec (\bomega ^g ) +  \bR ^g  \bh_{n-1}  (\btheta)  +  \frac{  \bA ^g} {\tau}\int_{n-2}^{\tau +n-2} \vec \( \bSigma_t \) dt  + \frac{ \bB ^g}{1-\tau}  \vec \( \br _{n-1} \br_{n-1}^{\top} \),  
			\end{eqnarray*}
		where $ \bQ_{n} =\(  (1-\tau)  \tau ^{-1}  \bvarrho_L \bB_H  + \bI_{r^2} \)\bQ_n ^H + \bQ_n ^L  $ ,
		$\bvarrho = \tau \bvarrho _H  +   (1-\tau) (\bvarrho_L\bR_H +  \bvarrho_L \bB_H \bvarrho_H) $,
		$\vec (\bomega^g) =  \(\bI_{r^2} - \bvarrho  \bR_L \bR_H \bvarrho  ^{-1} \)    [  \left \{ ( 1-\tau)  \bvarrho_L \bB_H +\tau \bI_{r^2} \right \} \{ 2  \bvarrho_{H3} \bR_H  \vec(\bomega_{H1}) -  \bvarrho_{H2}   \vec(\bomega_{H2}) $ $ + \(\bvarrho _{H2} - 2 \bvarrho_{H3}  \) \vec ( \bnu \bnu ^{\top}) \}  
		+ (1-\tau) \{ 2  \bvarrho_{L3} \bR_L  \vec(\bomega_{L1}) -  \bvarrho_{L2}   \vec(\bomega_{L2})   \} +  (1-\tau )  \bvarrho_L  \vec \(  \bomega_{H}    \) ] $ $  +  \bvarrho   \vec \(  \bomega_L \) + \bvarrho  \bR_L \vec \(\bomega_H \) $,
		$\bR ^g = \bvarrho   \bR_L \bR_H \bvarrho  ^{-1}$,
		$\bA ^g= \bvarrho   \bR_L \bB_H $,
		$\bB ^g = \bvarrho   \bB_L$.
	\end{itemize}
\end{thm}

Theorem \ref{Prop-IV} (c) shows that the daily integrated  factor volatility can be decomposed into $\bh_n (\btheta)$ and  $\bQ_{n}$,  
where $\bh_{n}(\btheta ) $ is adapted to the filtration, $\calF _{n-1}  = \sigma (  \bff (t), t \leq n-1 )$, and $\bQ_{n}$ is the martingale difference. 
Note that the conditional daily integrated factor volatility is in the famous VEC-GARCH(1,1) structure \citep{bollerslev1988capital} with the open-to-close integrated factor volatility ($\int_{n-2}^{\tau +n-2} \vec \( \bSigma_t \) dt$) and the close-to-open squared log factor return ($\vec \( \br _{n-1} \br_{n-1}^{\top} \)$) as innovations.
Similarly, $\bh_n^H(\btheta)$ and $\bh_n^L(\btheta)$ are conditional expected volatilities for the open-to-close and close-to-open periods, respectively, which are adapted to the filterations $\calF _{n-1}$ and $\calF _{\tau+n-1}$, respectively.  
They have the pervious open-to-close integrated factor volatility and the previous close-to-open squared log factor return  as innovations.
In Section \ref{SEC-Parameter}, we propose a parameter estimation procedure based on the inherent iterative structure in the integrated factor volatilities described in Theorem \ref{Prop-IV}. 
However, since we cannot observe the random fluctuation term $\bZ_t$ and instantaneous volatility process, the interceptor parameters $\bomega_{H1}$, $\bomega_{H2}$, $\bomega_{L1}$, $\bomega_{L2}$, and $\bnu$ are hard to be estimated directly. 
Thus, we develop a parameter estimation procedure for $\bomega_{H}^g$, $\bomega_{L}^g$, $\bgamma_H$, $\bgamma_L$, $\bbeta_H$, and $\bbeta_L$ instead. 
Moreover, to estimate the interceptor $\bomega^g$ for the whole-day dynamics,  we use the following expression:
	\begin{eqnarray*}
	\vec( \bomega^g )&=& \( \bI_{r^2} - \bR^g \)  \Big [ \left \{  (1-\tau) \bA_L^g + \tau \bI_{r^2}     \right \} \( \bI_{r^2} -\bR_H^g \) ^{-1} \vec(  \bomega_H^g) \cr
	&& \qquad \qquad \qquad \qquad \qquad \qquad + (1-\tau) \( \bI_{r^2} - \bR_L^g\)^{-1} \vec( \bomega_L^g)  \Big ].
	\end{eqnarray*}
If instantaneous volatility estimators are available, then we are able to estimate the interceptor parameters $\bomega_{H1}$, $\bomega_{H2}$, $\bomega_{L1}$, $\bomega_{L2}$. 
However, this is not the focus of the paper, so we leave it for a future study.

\begin{remark}
To reduce the model complexity, we can consider a scalar BEKK model \citep{caporin2008scalar, christoffersen2014correlation,  engle1995multivariate} as follows:
\begin{eqnarray*}
\bh_n (\btheta)  = \omega \vec (\bSigma ) + \gamma  \bh_{n-1}  (\btheta)  +  \frac{ a} {\tau}\int_{n-2}^{\tau +n-2} \vec \( \bSigma_t \) dt  + \frac{ b}{1-\tau}  \vec \( \br _{n-1} \br_{n-1}^{\top} \),  
\end{eqnarray*}
where $\bSigma$ is a long term average of  factor volatilities, and $a, b, \gamma$, and $\omega$ are non-negative scalar values such that $a+ b+ \gamma <1$.
This result can be obtained by choosing 
$\bomega_{H1}$, $\bomega_{H2}$, $\bomega_{L1}$, and $\bomega_{L2}$ as scaled $\bSigma$ and $\bnu$, $\bbeta_H$, $\bbeta_L$, $\bgamma_H$ and $\bgamma_L$ as scaled identity matrices.  
With an appropriate choice of scaled values, we can have $\omega = 1- a-b-\gamma$. 
 
\end{remark}

\begin{remark}
We can impose different dynamic models on the open-to-close and close-to-open periods. 
Then, the daily conditional expected integrated volatility is the sum of their conditional expected integrated volatilities. 
In this point of view, FOGI is the special example of the above general model. 
However, in practice, there is a trade-off between model error and estimation error.
Thus, it is important to develop a simple model that can capture the open-to-close and close-to-open dynamics. 
Under this parsimonious principle, we propose the FOGI model.
\end{remark}


\section{Estimation procedure}
\label{SEC-Parameter}

In this section, we present an estimation procedure for the model parameters and establish its asymptotic theorems.

\subsection{A model set-up}

We assume that the factor process of the true stock log prices follows the FOGI model in Definition \ref{Def-MGARCH-Ito}.
Due to the imperfections of the trading mechanisms, high-frequency data are contaminated by market microstructure noises so that the observed stock prices are in a noisy version of the true stock prices \citep{ait2009high}. 
In contrast, for low-frequency data, the effect of the microstructure noises is negligible, thus, we assume that the true log prices $\bX (t)$ and $\bX(t+\tau), t=0, \ldots, n$, at the market opening and closing times are observed.  
In light of this, we assume that the high-frequency intraday observations $\bX(t_{k,\ell}), \ell=1, \ldots, m-1,$ where $k-1=t_{k,0} < \cdots <  t_{k,m} = k-1+\tau$, are masked by the microstructure noises as follows:
	\begin{equation} 
	\label{eq-3.2}
	Y _{i}(t_{k, \ell}) = X_{i} ( t_{k, \ell})+\epsilon _{i}( t_{k, \ell}), \quad i=1,\ldots,p, k=1,\ldots,n, \ell =1, \ldots, m-1,
	\end{equation}
where the microstructure noises $\epsilon_{i}$'s are independent random variables with mean zero and variance $\eta_{ii}$. 
The noises are also independent of the price and volatility processes. 
For simplicity, we assume that the data are synchronized and the observed time points $t_{k, \ell}$'s are equally spaced, that is, 
$t_{k, \ell}- t_{k, \ell-1} = m ^{-1}$ for $k=1,\ldots, n$ and $\ell=1,\ldots, m$.

\begin{remark}
In practice, for the multivariate stock price process, high-frequency data are not synchronized, while observed time points are not equally spaced. 
This is the so-called non-synchronization problem and has been well studied in the finance literature. 
Data synchronization schemes include the refresh time \citep{barndorff2011multivariate}, the previous tick time \citep{wang2010vast, zhang2011estimating}, and the generalized sampling time \citep{ait2010high}. 
Under some mild conditions, these schemes provide the tools for solving the non-synchronization problem and to further obtain estimators for the daily integrated volatility given high-frequency data \citep{ait2010high, barndorff2011multivariate, bibinger2014estimating, christensen2010pre, wang2010vast, zhang2011estimating}.  
To highlight the modeling of the parametric volatility process, we make the simple assumption that the high-frequency observations are equally spaced and synchronized. 
\end{remark}

Given high-frequency data from the market open-to-close period, we can estimate the integrated volatility with the realized volatility matrix estimators, such as the multi-scale realized volatility matrix (MSRVM) \citep{zhang2011estimating}, the pre-averaging realized volatility matrix  (PRVM) \citep{christensen2010pre}, and the kernel realized volatility matrix (KRVM) \citep{barndorff2011multivariate}. 
See also \citet{ait2010high, barndorff2008designing, bibinger2014estimating, fan2007multi, fan2018robust, jacod2009microstructure, kim2018large, xiu2010quasi, zhang2006efficient, wang2010vast} for related research works. 
These non-parametric realized volatility matrix estimators can estimate historical volatility  well. 
Moreover, as the number of high-frequency observations $m$ goes to infinity, the majority of these non-parametric estimators can achieve the optimal convergence rate of $m^{-1/4}$ for estimating the open-to-close integrated volatility matrix in the presence of market microstructure noises.
In this paper, we denote such estimators by $\hat{\bGamma}_k^H$, $k=1,\ldots,n$, while in the numerical study, we use the PRVM estimator with weight function $\max(x, 1-x)$.


\subsection{Factor loading and volatility matrices} 
\label{SEC Factor}

The factor volatility process $\bff(t)$ is latent so that we are required to impose certain structures for its identification. 
Researchers often assume that the factor loading matrix $\bU$ is orthogonal while the factor volatility matrix $\bPsi_k$ (or $\bsigma_t^{\top} \bsigma_t$) is diagonal \citep{ait2015using, fan2013large, kim2019factor}. 
Under such assumptions, $\bU$ and $\bPsi_k$ correspond to the eigenmatrix and eigenvalues of the daily integrated volatility matrix $\bGamma_k$, respectively. 
In this case, our proposed FOGI model returns to the factor GARCH-It\^o model structure  introduced by \citet{kim2019factor}, as the factor GARCH-It\^o model describes the eigenvalue dynamics of the factor volatility process. 
Here the diagonal assumption on $\bPsi_k$ is rather restrictive in the sense that cross-sectional dynamics cannot be studied. 
Under such assumptions, the market dynamics only come from each factor marginally, while the correlations among factors are kept stable over time, which makes it difficult to explain the dynamics among factors. 
On the other hand, \citet{kim2018large} proposed the factor diffusion process to account for the grouping effects. 
Such grouping effects include sector and industry classifications, in which case one could use the global industry classification standard (GICS) as the group membership. 
Specifically, the factor loading matrix $\bU= (U_{ij}) _{i=1,\ldots, p, j =1,\ldots, r}$ is the membership matrix, where $U_{i,j} $ is equal to one if the $i$th asset belongs to the $j$th factor, and it is equal to zero, otherwise.
However, the performance of such models depends on the choice of the group membership. 
Recently, \citet{kim2022unified} suggested that one could estimate the factor loading matrix using the eigenmatrix of the variance from all daily integrated volatility matrices $\bGamma_{k}$'s, and this method does not require $\bPsi_k$ to be diagonal. 
Moreover, the factor loading matrix can be estimated non-parametrically under some stationary conditions, which, however, may be restrictive to model the idiosyncratic volatility matrix.
See also \citet{ait2015using, kong2017number, kong2018systematic, kong2021discrepancy,  pelger2019large}.

As discussed above, each approach has its own advantages and disadvantages, and the factor volatility matrix can be identified as long as the factor loading matrix $\bU$ can be well estimated. 
However, the estimation of $\bU$ is not the focus of this paper, and so we depend on existing research in the literature for this part and denote the consistent estimator of $\bU$ by $\hat{\bU}$.


\subsection{Weighted least squares estimation}
\label{SEC-QMLE}

In this section, we propose a quasi-maximum likelihood estimation procedure to estimate the true parameter $\btheta_0= (\vech(\bomega_{H0}^g), \vech(\bomega_{L0}^g), \vech(\bgamma_{H0}), \vech(\bgamma_{L0}), \vech(\bbeta_{H0}),\vech(\bbeta_{L0}), \bmu_{L0})$ of the FOGI model and establish its asymptotic behavior.
Let $d$ be the cardinality of $\btheta$, that is, $d = | \btheta|$. 
We note that $\rho$ does not have a role to define the conditional expected volatility matrix; thus, we do not include $\rho$ in the parameter of interest.

To estimate the model parameters, we examine the volatility dynamics during the open-to-close and close-to-open periods separately and, then, draw combined inferences. 
Specifically, the integrated volatilities hold the following relationships (see Theorem \ref{Prop-IV}):
	\begin{eqnarray*}
	&& \int_{n-1}^{\tau+n-1} \vec \( \bSigma_t  \) dt = \tau \bh_n^H (\btheta_0) + \bQ_{n}^H, \cr
	&& \int_{\tau+ n-1}^{n} \vec \( \bSigma_t  \) dt = (1- \tau) \bh_n^L (\btheta_0) + \bQ_{n}^L  \text{ a.s.}
	\end{eqnarray*}
For the open-to-close period, we can estimate the integrated factor volatility well with the MSRVM, the PRVM, or the KRVM. 
For the close-to-open period, high-frequency data are not available so that we approximate the integrated volatility with the close-to-open squared log factor return. 
Specifically, It\^o's lemma provides 
	\begin{equation}\label{eq-return}
	\vec \(\br_n(\bmu_{L0})  \br_n(\bmu_{L0}) ^{\top}\)= (1- \tau) \bh_n^L (\btheta_0) + \bQ_{n}^{LL}  \text{ a.s.},
	\end{equation}
where $\bQ_n^{LL} = \bQ_n^L + \( \int _{\tau+ n-1} ^{n}  r_{i,t}  d r_{j,t} +\int _{\tau+ n-1} ^{n}   r_{j,t}  d r_{i,t}\)_{i,j=1,\ldots,p}$ and  $\br_n(\bmu_L) = \bff (n) -\bff(\tau + n-1) -(1-\tau) \bmu_L$.
Thus, we can use the integrated factor volatility estimators and the close-to-open squared log returns $\br_n(\bmu_L)  \br_n(\bmu_L) ^{\top}$ as proxies for the open-to-close and close-to-open conditional volatilities, respectively. 
The accuracy of the proxies can be measured by the variance of the martingale differences $\bQ_n^H$ and $\bQ_n^{LL}$. 
To reflect the variance information on the estimation procedure, we adopt the famous weighted square loss function as follows:
	\begin{eqnarray*}
	L _{n} ( \btheta) &=& \frac{1} {2n}\sum _{k=1} ^{n} \Big [  \left \{ \vech(IV_k^f ) -  \tau \bh_{h,k}^H (\btheta ) \right \} ^{\top} \bV_H ^{-1}  \left \{ \vech(IV_k^f) -  \tau \bh_{h,k}^H (\btheta ) \right \}   \cr
	&&  \qquad \qquad +  \left \{ \br_{h,k}^2 (\bmu_L) - (1- \tau) \bh_{h,k}^L (\btheta ) \right \} ^{\top} \bV_L ^{-1}  \left \{\br_{h,k}^2 (\bmu_L) -  (1-\tau) \bh_{h,k}^L (\btheta ) \right \}  \Big ],
	\end{eqnarray*}
where $\bV_H$ and $\bV_L$ are covariance matrices for the $\bQ_n^{H}$ and $\bQ_n^{LL}$ terms, respectively. 
Moreover, 
$IV_k^f  = \int_{k-1}^{\tau+k-1} \bSigma_t dt $, 
$\bh_{h,n}^H (\btheta) = \vech \( \vec ^{-1} \(\bh_n^H (\theta) \) \)  $, $\bh_{h,n}^L (\btheta) = \vech \( \vec ^{-1} \(\bh_n^L (\theta) \) \)  $, and $\br_{h,k}^2 (\bmu_L) =\vech \( \br_k(\bmu_L)  \br_k(\bmu_L) ^{\top}\)$. 
Unfortunately, the latent factor process $\bff(t)$ is not observable. 
Thus, to implement the above weighted square loss function, we need to estimate the latent factor process.
As discussed in Section \ref{SEC Factor}, we impose some structures on the factor loading matrix $\bU$ so that it is estimable, which is the minimum requirement for investigating the latent factor model. 
See \citet{ait2015using, fan2013large, kim2018large, kim2019factor, kim2022unified} for more details.
With a consistent estimator $\hat{\bU}$, we now develop an estimation procedure for the model parameters in the high-dimensional set-up, where the number of assets are allowed to diverge, as the number of observations diverges. 
We first need to estimate the open-to-close factor volatility matrices and the close-to-open factor returns as follows:
	\begin{eqnarray*}
	RV_k  =p^{-2} \hat{\bU}^{\top}  \hat{\bGamma}_k ^H \hat{\bU}  \quad \text{and} \quad \hat{\br}_k   (\bmu_L) =  p^{-1} \hat{\bU} ^{\top} [\bX (k) -\bX(k-1+\tau )]  - (1-\tau) \bmu_L,
	\end{eqnarray*}
	where $\hat{\bGamma}_k^H$ is the realized volatility matrix estimator, such as the MSRVM, the PRVM, or the KRVM given the $k$th period open-to-close high-frequency data.
The conditional volatilities are 
	\begin{eqnarray*}
	&&\hat{\bh}_n^{H}(\btheta)  =  \vec \( \bomega_{H}^g\) + \bR_H^g \hat{  \bh}_{n-1} ^{H} (\btheta)  +  \frac{  \bA_H^g} {\tau} \vec \(RV_{n-1} \)   + \frac{ \bB_H^g}{1-\tau}  \vec \( \hat{\br} _{n-1} (\bmu_L)  \hat{\br}_{n-1}^{\top} (\bmu_L) \), \cr
	&&\hat{ \bh}_n^{L} (\btheta)  = \vec \( \bomega_{L}^g\) + \bR_L^g  \hat{\bh}_{n-1} ^{L} (\btheta)  +  \frac{  \bA_L^g} {\tau} \vec \(RV_{n} \)   + \frac{ \bB_L^g}{1-\tau}  \vec \( \hat{\br} _{n-1}  (\bmu_L)  \hat{\br}_{n-1}^{\top}(\bmu_L)  \).
	\end{eqnarray*}
Then, the weighted square loss function is
	\begin{eqnarray*} 
	\hat{L} _{n,m}  ( \btheta) &=& \frac{1} {2n}\sum _{k=1} ^{n} \Big [   \left \{ \vech(RV_k ) -  \tau \hat{\bh}_{h,k}^{H} (\btheta ) \right \} ^{\top} \hat{\bV}_H ^{-1}  \left \{ \vech(RV_k ) -  \tau \hat{\bh}_{h,k}^{H} (\btheta ) \right \}   \cr
	&&  \qquad   +   \left \{ \hat{\br}_{h,k}^{2} (\bmu_L) - (1- \tau) \hat{\bh}_{h,k}^{L} (\btheta ) \right \} ^{\top} \hat{\bV}_L ^{-1}  \left \{\hat{\br}_{h,k}^{2} (\bmu_L) -  (1-\tau) \hat{ \bh}_{h,k}^{L} (\btheta ) \right \}  \Big ],
	\end{eqnarray*}
where $\hat{\bh}_{h,n}^{H} (\btheta) = \vech \( \vec ^{-1} \(\hat{\bh}_n^{H} (\theta) \) \)$, $\hat{\bh}_{h,n}^{L} (\btheta) = \vech \( \vec ^{-1} \( \hat{\bh}_n^{L} (\theta) \) \)$,
and $\hat{\br}_{h,k}^{2} (\bmu_L) =\vech \( \hat{\br} _{n-1}  (\bmu_L)  \hat{\br}_{n-1}^{\top}(\bmu_L)  \)$.
We estimate the true parameter $\btheta_0$ by minimizing $\hat{L} _{n,m}  ( \btheta) $, that is,
	\begin{equation}
	\label{WLSE-Large}
	\hat{\btheta}  = \arg \min_{\btheta \in \bTheta} \hat{L}_{n,m}  (\btheta),
	\end{equation}
where $\bTheta$ is the parameter space of $\btheta$.
We choose the weight matrices $\hat{\bV}_H$ and $\hat{\bV}_L$ as discussed in Section \ref{SEC-weight}.
We call it the weighted least square estimator (WLSE).

To establish the asymptotic theorems for the proposed WLSE $\hat{\btheta}$, we require the following conditions.

\begin{assumption}\label{Assumption1}
	~
	\begin{enumerate}
		\item [(a)] $\bTheta$ is compact; $\bomega_H^g$ and $\bomega_L^g$ are positive definite; $\bbeta_H$, $\bbeta_L$, $\bgamma_H$ and $\bgamma_L$ are symmetric matrices; their (1,1)th elements are restricted to be positive; $\|\tilde{\bA} \|_2$,   $\|\bbeta_H\|_2$, $\|\bbeta_L\|_2$, $\|\bgamma_H\|_2$ and $\|\bgamma_L\|_2$ are less than 1, where   $\bzero $ is a $r^2$-by-$r^2$ zero matrix, and
		 \begin{equation*}
   \tilde{\bA} =     	\begin{pmatrix}
\bA^g _H & \frac{\tau \bB_H^g}{1-\tau} & \tau \bR^g_H & \bzero   \\ 
\frac{(1-\tau) \bA^g _L}{\tau} & \bB_L^g & \bzero &  (1-\tau)\bR^g_L    \\ 
\frac{\bA^g _H}{\tau} & \frac{\bB_H^g}{1-\tau} & \bR^g_H & \bzero     \\ 
\frac{ \bA^g _L}{\tau} & \frac{\bB_L^g}{1-\tau} & \bzero &  \bR^g_L  
\end{pmatrix} ;
    \end{equation*}
     $\btheta_0$ is an interior point of $\bTheta$; 
		

		\item  [(b)] $ \max_{k}  E \[ \left \| 
 \hat{\bGamma}_k^H - IV _k  \right \|_{F}^4 \]  = O(p^4 m^{-1})$, where $IV _k$ is the integrated volatility for the open-to-close period of the $k$th day; 
		
		\item [(c)]  $ \hat{\bV}_H$ and  $  \hat{\bV}_L$ are strictly positive definite and  consistent estimators for $\bV_H$ and $\bV_L$;
		
		\item [(d)]
		There exists a factor loading matrix estimator $\hat{\bU}= (\hat{U}_{ij})_{i=1,\ldots, p, j =1,\ldots, r}$ such that, for some $b \geq 4$, 
		$$
		\max_{1 \leq i \leq r} \E \( \| p^{-1/2} \hat{\bU}_i - p^{-1/2} sign (\hat{\bU}_i ^{\top} \bU_i)  \bU_i  \|_{F}^b \) \leq C  \nu_{m,n} ^b,
		$$
		where $\bU_i$ and $\hat{\bU}_i$ are the $i$th column of $\bU$ and $\hat{\bU}$, respectively, and $\nu_{m,n} = o(1)$;

		\item [(e)] $U_{ij} =O(1)$ and $\bU^{\top} \bU = p \bI_r$;
		
		\item [(f)]  $\pi(p) / p =o(1)$.
	\end{enumerate}
\end{assumption}

\begin{remark}
Under the finite fourth moment condition, \citet{kim2016asymptotic} showed that the realized volatility estimators satisfy Assumption \ref{Assumption1}(b). 
Under some stationary condition, we can show  Assumption \ref{Assumption1}(c) (see Section \ref{SEC-weight}).
\end{remark}
\begin{remark}
When $p^{-1/2}\bU$ is the eigenmatrix of the factor volatility matrix,  Theorem 3.1 from \cite{kim2019factor} shows that $\nu_{m,n}$ is $m^{-1/4} +n^{-1/2}+ \pi(p)/p$ in Assumption \ref{Assumption1}(d).
If we assume that the factor loading matrix is known, such as the GICS membership matrix, $\nu_{m,n}$ will be zero. 
When the membership matrix is unknown and we provide its estimator, the convergence rate $\nu_{m,n}$ will be the corresponding convergence rate of the estimator. 
\end{remark}

The following theorem establishes the asymptotic results for the proposed WLSE $\hat{\btheta} $.

\begin{thm}\label{Thm:Theta-Large}
Under Assumption \ref{Assumption1},  we have
	\begin{equation*}
	\left \| \hat{\btheta}   -\btheta_0 \right \|_{\max}  =   O_p\(  m^{-1/4} + n^{-1/2} +\nu_{m,n}+ (\pi(p)/p)^{1/2} \).
	\end{equation*}
\end{thm}

\begin{remark}
Theorem \ref{Thm:Theta-Large} shows that the proposed WLSE $\hat{\btheta} $ has the convergence rate $m^{-1/4}+n^{-1/2} +\nu_{m,n}+ (\pi(p)/p)^{1/2}$. 
The term $m^{-1/4}$ is coming from estimating the integrated volatility matrix $\bGamma_k$ non-parametrically, which is known as the optimal rate, given the presence of the market microstructure noises. 
The term $n^{-1/2}$ is the usual convergence rate of low-frequency analysis.
The terms $\nu_{m,n}$ and $(\pi(p)/p)^{1/2}$ are the cost to handle the large volatiltiy matrix. 
Specifically, the term $\nu_{m,n}$ is the cost to estimate the latent factor loading matrix $\bU$.
Since we employ the high-dimensional latent factor process, we have the extra term $(\pi(p)/p)^{1/2}$ for identifying the factor process. 
\end{remark}

\begin{remark}
To evaluate the proposed FOGI estimation procedure, we first need to estimate the number, $r$, of factors. 
Usually, the number of factors is estimated based on eigenvalues of sample covariance estimators \citep{alessi2010improved, ahn2013eigenvalue, bai2002determining, choi2018multilevel, giglio2021asset, onatski2010determining, trapani2018randomized}.
In high-frequency finance, we can use realized volatility matrix estimators instead of sample covariance estimators. 
In this paper, we have high-frequency and low-frequency observations; thus, we can use the following averaged open-to-close and close-to-open volatility estimators: 
\begin{align*}
	\hat{\bar{\bGamma}}^{H}_n &=  \frac{1}{n} \sum_{k=1}^n \hat{\bGamma}_k ^H, \cr
	\hat{\bar{\bGamma}}^{L}_n &= \frac{1}{n} \sum_{k=1}^{n} (\bX(k) -\bX(\tau+k-1) -\bar{\bX} )  (\bX(k) -\bX(\tau+k-1)  -\bar{\bX} ) ^{\top},
\end{align*}
\begin{equation*}
	\hat{r} = \arg \underset{1 \leq j \leq r_{\max}}{\min} \[ K(\hat{\bar{\bGamma}}^{H}_n, n, m, j) + K(\hat{\bar{\bGamma}}^{L}_n, n, 1, j) \],
\end{equation*}
where
\begin{equation*}
  K(\bGamma,n,m,j) = p^{-1}  \lambda_{j} + j \times c_1 \left\lbrace \sqrt{\log p / (nm^{1/2})} + p^{-1} \log p \right\rbrace^{c_2}  -1
  ,
\end{equation*}
$\lambda_{j}$ is the $j$th largest eigenvalue of $\bGamma$.
Under the pervasive condition (Assumption \ref{Assumption-Condition}(a)), we can show its consistency similar to the proofs of  \cite{ait2015using}. 
\end{remark}


\subsection{Choice of weight matrices for the WLSE} 
\label{SEC-weight}

The performance of the proposed WLSE procedure depends on the choice of the weight matrices $\bV_H$ and $\bV_L$.
To reflect the accuracy of the information of the data sources, we choose the weight matrices $\bV_H$ and $\bV_L$  as the variances of the martingale difference terms.
For example, for the open-to-close period, the difference between the GARCH volatility $\tau \bh_{h,k}^H (\btheta_0)$ and the proxy $\vech(IV_k^f) $ is $\bQ_{h,k}^H$, while, for the close-to-open period, the difference between the GARCH volatility $  (1- \tau) \bh_{h,k}^L (\btheta_0 ) $ and the proxy $\vech \( \br _{k-1} \br_{k-1}^{\top} \)$ is $ \bQ_{h,k}^{LL} $.
Then $\bV_H$ and $\bV_L$ are the covariance matrices of the martingale differences $\bQ_{h,k}^H$ and $\bQ_{h,k}^L$, respectively.
To estimate the variance of the martingale difference terms, we first estimate the GARCH volatilities for the open-to-close and close-to-open periods separately.
Specifically, we estimate the GARCH parameters $\btheta^g_{H,0}= (\vech \( \bomega_{H0}^g\) , \vec (\bR_{H0}^g), \vec(  \bA_{H0}^g), \vec( \bB_{H0}^g), \bmu_{L0} ) $ and $\btheta^g_{L,0}= (\vech \( \bomega_{L0}^g\) , \vec (\bR_{L0}^g), \vec(  \bA_{L0}^g),$ $\vec( \bB_{L0}^g), \bmu_{L0} )$ by
	\begin{equation*}
	\hat{\btheta}_H^g = \arg \min_{\btheta_H^g}  \frac{1} {2n}\sum _{k=1} ^{n}     \left \{ \vech(RV_k) -  \tau \hat{\bh}_{h,k}^H (\btheta) \right \} ^{\top}   \left \{ \vech(RV_k) -  \tau \hat{\bh}_{h,k}^H (\btheta) \right \} 
	\end{equation*}
and 
	\begin{equation*}
	\hat{\btheta}_L^g = \arg \min_{\btheta_L^g}    \frac{1} {2n}\sum _{k=1} ^{n}     \left \{ \hat{\br}_{h,k}^2 (\bmu_L) - (1- \tau) \hat{\bh}_{h,k}^L (\btheta) \right \} ^{\top}    \left \{\hat{\br}_{h,k}^2 (\bmu_L) -  (1-\tau) \hat{ \bh}_{h,k}^L (\btheta) \right \}.
	\end{equation*}
Similar to Theorem \ref{Thm:Theta-Large}, we can show that the least squares estimators (LSE) $\hat{\btheta}_H^g $ and $\hat{\btheta}_L^g$ are consistent with the convergence rate $  m^{-1/4} + n^{-1/2} +\nu_{m,n}+ (\pi(p)/p)^{1/2}$ under Assumption \ref{Assumption1}.  
Then, with these consistent estimators, we calculate their sample covariances as follows:
	\begin{eqnarray} 
	\label{eq-cov}
	&&\tilde{\bV}_H = \frac{1}{n} \sum_{k=1}^n   \left \{ \vech(RV_k) -  \tau \hat{\bh}_{h,k}^H (\hat{\btheta}_H^g ) \right \}   \left \{ \vech(RV_k) -  \tau \hat{\bh}_{h,k}^H (\hat{\btheta}_H^g ) \right \}   ^{\top}  , \cr
	&&\tilde{\bV}_L = \frac{1}{n} \sum_{k=1}^n   \left \{\hat{\br}_{h,k}^2 (\hat{\bmu}_L) - (1- \tau) \hat{\bh}_{h,k}^L (\hat{\btheta}_L^g ) \right \}   \left \{ \hat{\br}_{h,k}^2 (\hat{\bmu}_L) - (1- \tau) \hat{\bh}_{h,k}^L (\hat{\btheta}_L^g) \right \}   ^{\top}  .
	\end{eqnarray}
Under the stationary condition, we can show that $\tilde{\bV}_H$ and $\tilde{\bV}_L$ converge to their corresponding covariance matrices using the martingale convergence theorem. 
Due to the estimation errors from the finte sample estimation, some eigenvalues of the estimated weight matrices can be too small, which makes the estimators of the inverse weight matrices unstable. 
We therefore added a small positive value to the eigenvalues of the weight matrices.
Specifically, we employed consistent estimators of $\bV_H$ and $\bV_L$ as follows:
\begin{equation*}
	\hat{\bV}_H = \tilde{\bV}_H + \upsilon_n^H \bI_{r(r-1)/2} \qquad \text{and} \qquad \hat{\bV}_L = \tilde{\bV}_L + \upsilon_n^L \bI_{r(r-1)/2},
\end{equation*}
where $\tilde{\bV}_H$ and $\tilde{\bV}_L$ are defined in \eqref{eq-cov}, and $\upsilon_n^H$ and $\upsilon_n^L$ are tuning parameters, converging to zero.
In the numerical study, we choose $ \upsilon_n^H$ and $\upsilon_n^L$ as follows: 
$$
\upsilon_n^H = \frac{1}{r(r-1)/2}\sum_{i=1}^{r(r-1)/2} [\tilde{\bV}_H]_{ii} n^{-3/5} \quad \text{and} \quad \upsilon_n^L = \frac{1}{r(r-1)/2}\sum_{i=1}^{r(r-1)/2} [\tilde{\bV}_L]_{ii} n^{-3/5},
$$
where $\[\bM\]_{ij}$ denotes the $(i,j)th$ entry of a matrix $\bM$.
We note that to make the perturbation terms negligible, we should choose the rate faster than $n^{-1/2}$.
In contrast, to obtain stable inverse results, we need to choose relatively slow rate. 
In the numerical study, the choice of $n^{-3/5}$ shows a good performance, so we suggest $n^{-3/5}$ order.  
Thanks to this choice,  $\hat{\bV}_H$ and $\hat{\bV}_L$ are consistent estimators of $\bV_H$ and $\bV_H$, which have the same convergence rate as that of $\tilde{\bV}_H$ and $\tilde{\bV}_H$, respectively, so that all of the asymptotic properties  hold with $\hat{\bV}_H$ and $\hat{\bV}_L$.

We use the estimated weight matrices  $\hat{\bV}_H$ and $\hat{\bV}_L$ to evaluate the weighted square loss function $\hat{L} _{n,m} ( \btheta)$ in \eqref{WLSE-Large}. 
This procedure helps to assign more weights to the more informative and accurate proxy. 
Intuitively, the open-to-close period should be more informative given the abundant high-frequency observations, thus, more weight will be assigned to the open-to-close proxy. 
In the empirical study, $\hat{\bV}_H$ is smaller than $\hat{\bV}_L$ in terms of eigenvalues. 
Finally, with the consistent estimators $\hat{\bV}_H$ and $\hat{\bV}_L$, we can establish the asymptotic results in Theorem \ref{Thm:Theta-Large}.


\section{Large volatility matrix prediction} \label{SEC-Large}

\subsection{Idiosyncratic volatility matrices} 
\label{SEC-Idiosyncratic}

Recall that we impose the sparse condition on the idiosyncratic volatility matrix $\bGamma_{k}^s = \(\Gamma_{k, ij} ^s \) _{i,j=1,\ldots, p}$ as follows:
	\begin{equation*} 
 	\max_{1\leq j \leq p} \sum_{i=1}^p |\Gamma_{k, ij}^s| ^{\delta} |\Gamma_{k, ii}^s \Gamma_{k, jj}^s | ^{(1-\delta)/2} \leq M \pi(p) \quad \text{ a.s.},
	\end{equation*}
where $\delta \in [0,1)$, $M$ is the positive bounded random variable, and the sparsity level $\pi(p)$ diverges very slowly, such as $\log p$.
Furthermore, we assume  that the idiosyncratic volatility matrix  $\bGamma_{k}^s$ satisfies
	\begin{equation}
	\label{constant}
	\bGamma_k ^s = \bGamma^s  \;  a.s. \quad  \text{ for all } k=1,\ldots,n.
	\end{equation}
The constant condition \eqref{constant} is imposed so that the volatility matrices for the open-to-close period can be estimated consistently.

To estimate the sparse idiosyncratic volatility matrix $\bGamma ^s$, we follow the POET procedure introduced by \citet{fan2013large}.
Specifically, the input integrated volatility matrix estimator is estimated by
	\begin{equation*}
	\hat{\bar{\bGamma}}_n =  \frac{1}{n} \sum_{k=1}^n \hat{\bGamma}_k ^H + \frac{1}{n} \sum_{k=1}^{n} (\bX(k) -\bX(\tau+k-1) -\bar{\bX} )  (\bX(k) -\bX(\tau+k-1)  -\bar{\bX} ) ^{\top},
	\end{equation*}
where $\hat{\bGamma}_k^H$ is the realized volatility matrix estimator, such as the MSRVM, the PRVM, or the KRVM given the $k$th period open-to-close high-frequency data, and $\bar{\bX}$ is the sample mean of all close-to-open log returns.
We note that, under some regularity conditions, the input estimator  $\hat{\bar{\bGamma}}_n$ converges to $\bar{\bGamma}_n =\sum_{k=1}^n \bGamma_k/n$.
Then, $\hat{\bar{\bGamma}}_n$ admits
	\begin{equation*}
	\hat{\bar{\bGamma}}_n= \sum_{i=1}^p \hat{\lambda}_{i} \hat{\bq}_{i} \hat{\bq}_{i}^{\top},
	\end{equation*}
where $\hat{\lambda}_{i}$ is the $i$th largest eigenvalue of $\hat{\bar{\bGamma}}_n$ and $\hat{\bq}_{i}$ is its corresponding eigenvector.
The input idiosyncratic volatility matrix estimator is 
	\begin{equation*}
	\tilde{\bGamma}^s = (\tilde{\Gamma}_{ij}^s ) _{i,j=1,\ldots,p} = \hat{\bar{\bGamma}}_n -  \sum_{i=1}^r \hat{\lambda}_{i} \hat{\bq}_{i} \hat{\bq}_{i}^{\top},	
	\end{equation*}
and we apply the following adaptive thresholding scheme to the above input: 
	\begin{equation} \label{Sparse-est}
	\hat{\Gamma}_{ij}^s =
	\begin{cases}
	\tilde{\Gamma}_{ij}^s \vee 0, & \text{ if } i= j\\
	s_{ij} ( \tilde{\Gamma}_{ ij}^s) \1 ( |\tilde{\Gamma}_{ij}^s| \geq \varpi_{ij} ) ,  & \text{ if } i \neq j
	\end{cases}
	\quad \text{and} \quad \hat{\bGamma}^s = (\hat{\Gamma}_{ij}^s )_{1\leq i,j \leq p},
	\end{equation}
where the thresholding function $s_{ij} (\cdot) $ satisfies  $|s_{ij} (x) -x |\leq \varpi_{ij}$, and takes the thresholding level $\varpi_{ij} = \varpi_{m,n}\, \sqrt{ ( \tilde{\Gamma}_{ii}^s \vee 0 ) ( \tilde{\Gamma}_{jj}^s \vee 0 )} $ that is based on the correlation structure. 
The thresholding function includes interesting examples such as hard thresholding and soft thresholding functions.

\subsection{Large volatility matrix prediction}

In this section, we demonstrate how to predict future large volatility matrices and investigate the asymptotic behaviors of the proposed method. 
Recall that the daily integrated volatility matrix can be decomposed into the factor and idiosyncratic volatility matrices. 
Let the forecast origin be $n$, and we are interested in constructing an estimator for
	\begin{equation*}
	\E \( \bGamma_{n+1} \middle | \FF_n \)= \bU \bH_{n+1}(\btheta_0)  \bU^{\top} + \E \( \bGamma_{n+1} ^s\middle | \FF_n \) \quad  \text{ a.s.}
	\end{equation*}
For the latent factor volatility matrix $\bPsi_{n+1}$, the FOGI model structure produces the following iterative relationship
	\begin{eqnarray*}
	\bh_{n+1} (\btheta_0) 
	&=& \E \( \vec \( \bPsi_{n+1} \) \middle | \FF_n  \)  \cr
	&=& \vec (\bomega ^g_0 ) +  \bR_0 ^g  \bh_{n} ^H (\btheta_0)  +  \frac{  \bA ^g_0} {\tau}\int_{n-1}^{\tau +n-1} \vec \(\bsigma_t^{\top} \bsigma_t\) dt  + \frac{ \bB_0 ^g}{1-\tau}  \vec \( \br _{n} \br_{n}^{\top} \),  
	\end{eqnarray*}
where $\bomega ^g_0 $, $ \bR_0 ^g $, $\bA ^g_0$, and $\bB_0 ^g$  are defined in Theorem \ref{Prop-IV}, and $\vec(\bH_{n+1} (\btheta_0) )=\bh_{n+1} (\btheta_0)$. 
With the WLSE $\hat{\btheta} $ defined in \eqref{WLSE-Large}, we estimate the latent factor volatility matrix $\bH_{n+1}(\btheta_0)$ by
	\begin{equation}
	\label{Factor-est}
	\hat{\bH}_{n+1} (\hat{\btheta} ) = \vec ^{-1} \( \hat{ \bh}_{n+1} (\hat{\btheta} ) \) , 
	\end{equation}
where 
	\begin{equation*}
	\hat{ \bh}_{n+1} (\hbtheta )  = \vec \( \bomega^g\) + \bR^g  \hat{\bh}_{n} (\hbtheta )  +  \frac{  \bA^g} {\tau} \vec \(RV_{n} \)   + \frac{ \bB^g}{1-\tau}  \vec \( \br _{n}  (\bmu_L)  \br_{n}^{\top}(\bmu_L)  \).
	\end{equation*}
On the other hand, we impose the constant assumption on the idiosyncratic volatility matrix, that is, 
	\begin{equation*}
	\E \( \bGamma_{n+1}^s \middle | \FF_{n} \) = \bGamma^s \quad  \text{ a.s.}
	\end{equation*}
and estimate $\bGamma ^s$ using the POET procedure as discussed in Section \ref{SEC-Idiosyncratic}.
Combing the factor volatility matrix estimator in \eqref{Factor-est} and the idiosyncratic volatility matrix estimator in \eqref{Sparse-est}, we estimate the conditional expectation of the future large volatility matrix $\E \( \bGamma_{n+1} \middle | \FF_{n} \)$ by 
	\begin{equation*}
	\tilde{\bGamma}_{n+1} = \hat{\bU} \hat{\bH}_{n+1} (\hat{\btheta} ) \hat{\bU}^{\top} + \hat{\bGamma}^s. 
	\end{equation*}
Under some conditions, \citet{fan2018robust} showed that the idiosyncratic volatility matrix estimator is consistent in the sense of the matrix spectral norm, and they established the asymptotic theorems for the estimator of the future large volatility matrix under the factor GARCH-It\^o model. 
We employ their results to investigate the asymptotic behaviors of $\tilde{\bGamma}_{n+1}$ and require the following technical conditions.    

\begin{assumption}
	\label{Assumption-Condition}
	~
	\begin{enumerate}
		\item[(a)] There exists some fixed positive constant $c_1$ such that $\lambda_{n,1} / D_{\lambda} \leq c_1$ a.s., where $\lambda_{n,i}$ is the $i$th largest eigenvalue of $\bU \bPsi_n \bU ^{\top}$ and $D_{\lambda} =\min \{\lambda_{j,i} - \lambda_{j, i+1}: i=1,\ldots,r , j=1,\ldots, n\}$, and the smallest eigenvalue of $\bGamma_n^s$ stays away from zero;
		
		\item [(b)] For some fixed constant $c_2$, we have
		\begin{equation*}
		\frac{p}{r} \max_{ 1\leq  i \leq p} \sum_{j=1}^r q _{ij} ^2  \leq c_2 \text{ a.s.},
		\end{equation*}
		where $\bq_j = (q_{1j}, \ldots, q_{pj})^\top$ is the $j$th eigenvector of $\bU \bPsi_n \bU ^{\top}$;
		
		\item [(c)] The input volatility matrix $\hat{\bar{\bGamma}}_n$ satisfies 
		\begin{equation}\label{condition-Thm-future}
		\Pr \left \{\max_{1 \leq i,j \leq p} |\hat{\bar{\Gamma}}_{n, ij} -\bar{\Gamma}_{n, ij} | \geq C \sqrt{\frac{\log p}{m^{1/2} n +m}}  \right \} \leq p^{-1};
		\end{equation}
		
		\item[(d)] $\pi(p)/ \sqrt{p}  +\sqrt{    \log p  /  (m^{1/2} n +m) } =o(1).$
	\end{enumerate}
\end{assumption}

\begin{remark}
Assumption \ref{Assumption-Condition}(a)--(b) are called the pervasive condition and incoherence condition, respectively, which are widely employed to investigate the behavior of the low-rank matrix inferences \citep{candes2011robust, fan2016l1}.   
For example, the pervasive condition helps to identify the latent factor volatility matrix as the eigenvalues diverge with the order $p$ while the eigenvalues of the idiosyncratic volatility matrix are bounded. 
On the other hand, the incoherence condition makes it possible to establish the element-wise convergence rate of the factor volatility matrix. 
The sub-Gaussian concentration inequality condition Assumption \ref{Assumption-Condition}(c) is required to investigate the high-dimensional statistics. 
Under some regularity conditions and the martingale property of daily realized volatility estimators, the average realized volatility estimators can obtain this bound \citep{fan2018robust, tao2013optimal}. 
\end{remark}

\begin{thm} 
\label{Thm-LargePredict}
Under the assumptions of Theorem \ref{Thm:Theta-Large}, the sparsity condition \eqref{sparsity} and Assumption \ref{Assumption-Condition} are met. 
Take the thresholding level to be $ \varpi_{m,n}=C_{\varpi} b_{m,n}$ for some large fixed constant $C_{\varpi}$, where $b_{m,n} =  \pi(p) / p + \sqrt{ \log (p \vee m) /(nm^{1/2}+m)}$.
Then we have
	\begin{eqnarray}
	&& \| \hat{\bGamma} ^s  - \bGamma ^s  \|_{\max} = O_p \(   b_{m,n} \) , \label{Thm-LargePredict:result-1}\\
	&& \| \hat{\bGamma} ^s  - \bGamma ^s  \|_{2} = O_p\(  \pi(p) b_{m,n}^{1-\delta} \), \label{Thm-LargePredict:result-2}\\
	&&\| \tilde{\bGamma}_{n+1} - \E\( \bGamma_{n+1} | \FF_{n} \) \| _{\bGamma^* } = O_p  \Big (  m^{-1/4} + n^{-1/2} + \nu_{m,n} +   \frac{\pi(p)}{ p^{1/2}} \cr
	&& \quad \qquad \qquad \qquad \qquad \qquad \qquad \qquad +p^{1/2} (m^{-1/2} + n^{-1} + \nu_{m,n}^2)+ \pi(p)  b_{m,n}^{1-\delta} \Big ),  \label{Thm-LargePredict:result-3}
	\end{eqnarray}
where the relative Frobenius norm is defined as $\| \bM\|_{\bGamma^* }^2= p^{-1} \| \bGamma  ^{*-1/2} \bM \bGamma ^{* -1/2}\|_F ^2 $ where $\bGamma ^{*} = \E\( \bGamma_{n+1} | \FF_{n} \)$.
\end{thm}

\begin{remark}
Theorem \ref{Thm-LargePredict} shows that the estimator of the future volatility matrix $\tilde{\bGamma}_{n+1}$ has the convergence rate $m^{-1/4} + n^{-1/2} + \nu_{m,n} +   \frac{\pi(p)}{ p^{1/2}}   +p^{1/2} (m^{-1/2} + n^{-1} + \nu_{m,n}^2)+ \pi(p)  b_{m,n}^{1-\delta}$ under the relative Frobenius norm.
When the factor loading matrix $\bU$ is related to the eigenmatrix \citep{kim2019factor}, $\nu_{m,n}$ will be $m^{-1/4} + \pi(p)/p$. 
Then $\tilde{\bGamma}_{n+1}$ will be consistent, as long as $p=o(m + n^2)$.
That is, the number, $p$, of assets has slower order than the numbre, $m$, of hihg-frequency observations and the squared sample period $n^2$.   
\end{remark}



\section{Simulation study} 
\label{SEC-Simulation}

In this section, we show that the proposed methodology has a good finite sample performance in estimation and prediction via a simulation study. 
Let $p$ be the total number of assets, $n$ be the total number of low-frequency periods, and $m$ be the total number of high-frequency returns during each low-frequency period.
Let $r$ be the number of common market factors in the high-dimensional set-up. 
The number of daily trading hours in the US market is 6.5,  hence we choose $\tau=6.5/24$.

We now investigate the high-dimensional case for $p=200$ and $r=3$.  
For the instantaneous latent factor volatility process described by \eqref{def : UMGARCH}, we took the model parameters as follows:
\begin{eqnarray*}
	&&\vech(\bomega_{H1})=(0.06,0,0,0.08,0,0.04), \qquad \vech(\bomega_{H2})=(0.004,0,0,0.004,0,0.004),\cr 
	&&\vech(\bomega_{L1})=(0.024,0,0,0.012,0,0.004), \qquad \vech(\bomega_{L2})=(0.006,0,0,0.004,0,0.012), \cr 
	&& \vech(\bgamma_H)=(0.5,0,0,0.3,0,0.4), \qquad \vech(\bgamma_L)=(0.6,0,0,0.8,0,0.7), \cr 
	&& \vech(\bbeta_H)=(0.7,0,0,0.6,0,0.8), \qquad \vech(\bbeta_L)=(0.3,0,0,0.25,0,0.2), \cr
	&& \bnu = \diag(0.06,0.04,0.03), \qquad \bmu_H = \bmu_L = 0.
\end{eqnarray*}
It follows that the true parameter $\btheta_0$ is
\begin{equation*}
	\begin{split}
	\btheta_0 =& (\vech(\bomega_{H0}^g), \vech(\bomega_{L0}^g), \vech(\bgamma_{H0}), \vech(\bgamma_{L0}), \vech(\bbeta_{H0}),\vech(\bbeta_{L0}), \bmu_0) \\
	=& (0.0089, 0, 0, 0.0045, 0, 0.0018, 0.0075, 0, 0, 0.0031, 0, 0.0018, 0.5, 0, 0, 0.3, 0, 0.4,\\
	&  0.6, 0, 0, 0.8, 0, 0.7, 0.7, 0, 0, 0.6, 0, 0.8, 0.3, 0, 0, 0.25, 0, 0.2,0,0,0).
	\end{split}
\end{equation*} 
Initial values for simulation were chosen to be 
\begin{equation*}
	\begin{split}
	&\vech(\bSigma_0)=\vech(\bomega^g)=c(0.0100, 0, 0, 0.0044, 0, 0.0024) , \\
	&\bff(t)=(10,10,10)^\top , \qquad  \bX(t) = (10, \ldots, 10)^\top.
	\end{split}
\end{equation*}
For the factor loading matrix $\bU$, $\bU_1$ takes values $\sqrt{2} \cos \left( 2i \pi/p \right)$, $i=1, \ldots, p$, $\bU_2$ entries share the same value $1$, and $\bU_3$ takes values $\sqrt{2} \sin \left( 2i \pi/p \right)$, $i=1, \ldots, p$. 
Then the factor loading matrix retains the structure such that $\bU^{\top} \bU=p \bI_r$, which is the condition required in \citet{kim2022unified} for obtaining a consistent estimator of $\bU$ by using the eigenmatrix of the variance of all daily integrated volatility matrices. 
To generate the sparse idiosyncratic diffusion process in its daily integrated co-volatility $\bGamma^s=\left(\Gamma^s_{ij} \right)_{1 \leq i,j \leq p}$, we took  
	\begin{equation*}
	\Gamma^s_{ij} = 0.5^{ |i-j| } \sqrt{\Gamma_{ii}^s \Gamma_{jj}^s}
	\end{equation*}
for the off-diagonal elements and $\Gamma^s_{ii}=0.004, i=1, \ldots, p$, for the diagonal elements. 
Market microstructure noises, $\epsilon _{i}( t_{k, \ell})$, were modeled by independent $0.005 N(0,\Gamma_{k,ii})$, for $i=1,\ldots,p$, $k=1,\ldots,n$, and $\ell =1, \ldots, m-1$. 
Given high-frequency data, the PRVM estimator was again adopted to obtain the daily large realized volatility matrix estimates $\hat{\bGamma}^H_k$, $k=1,\ldots,n$. 
To estimate the factor loading matrix, we followed the procedure given in \citet{kim2022unified}. 
Specifically, let $\hat{\bar{\bGamma}}^H_n=\sum_{k=1}^n \hat{\bGamma}^H_k/n$ be the sample mean and $\hat{\bS}_n^H=\sum_{k=1}^n (\hat{\bGamma}^H_k-\hat{\bar{\bGamma}}^H_n)^2/(np)$ be the sample variance. 
We estimated $p^{-1/2} \bU$ by the first $r$ eigenvectors of $\hat{\bS}_n$ and computed the factor volatility matrix estimates $RV_k $ by $RV_k =p^{-2} \hat{\bU}^\top \hat{\bGamma}^H_k \hat{\bU}$.
We note that to estimate the eigenvectors, we only utilized the open-to-close high-frequency data
 because the daily integrated volatility matrices should be consistently estimated to estimate the variance of volatility matrices, $\hat{\bS}^{H}_{n}$, well.
 Thus, under the assumption that the factor loadings are constant over time, we can estimate the eigenvectors consistently by only using the high-frequency observations.  
 In contrast, we utilized low-frequency data for idiosyncratic volatility matrices as described in Section \ref{SEC-Idiosyncratic}  because, technically, we do not need consistent daily integrated volatility matrix estimators and assume that the idiosyncratic volatility matrix is constant over time.
The constant assumption helps reduce the complexity of the model and enjoy the benefit of incorporating low-frequency data for the estimation of the idiosyncratic volatility matrix.
 The close-to-open factor returns $r _k(\mathbf{0})=p^{-1} \hat{\bU}^\top [\bX(k)-\bX(k-1+\tau)]$ were obtained.
We employed $\tilde{\bV} _{H}$, and $\tilde{\bV} _{L}$ defined in \eqref{eq-cov} for the consistent estimators of $\bV _{H}$, and $\bV _{L}$, respectively. 
Parameter estimates were obtained by minimizing the proposed weighted square loss function $\hat{L}_{n,m}(\btheta)$. 
We took $n=125,250,500$ and $m=390,780,2340$. 
We repeated the simulation 200 times.

\begin{figure}[t!]
\centering
\includegraphics[width=\textwidth]{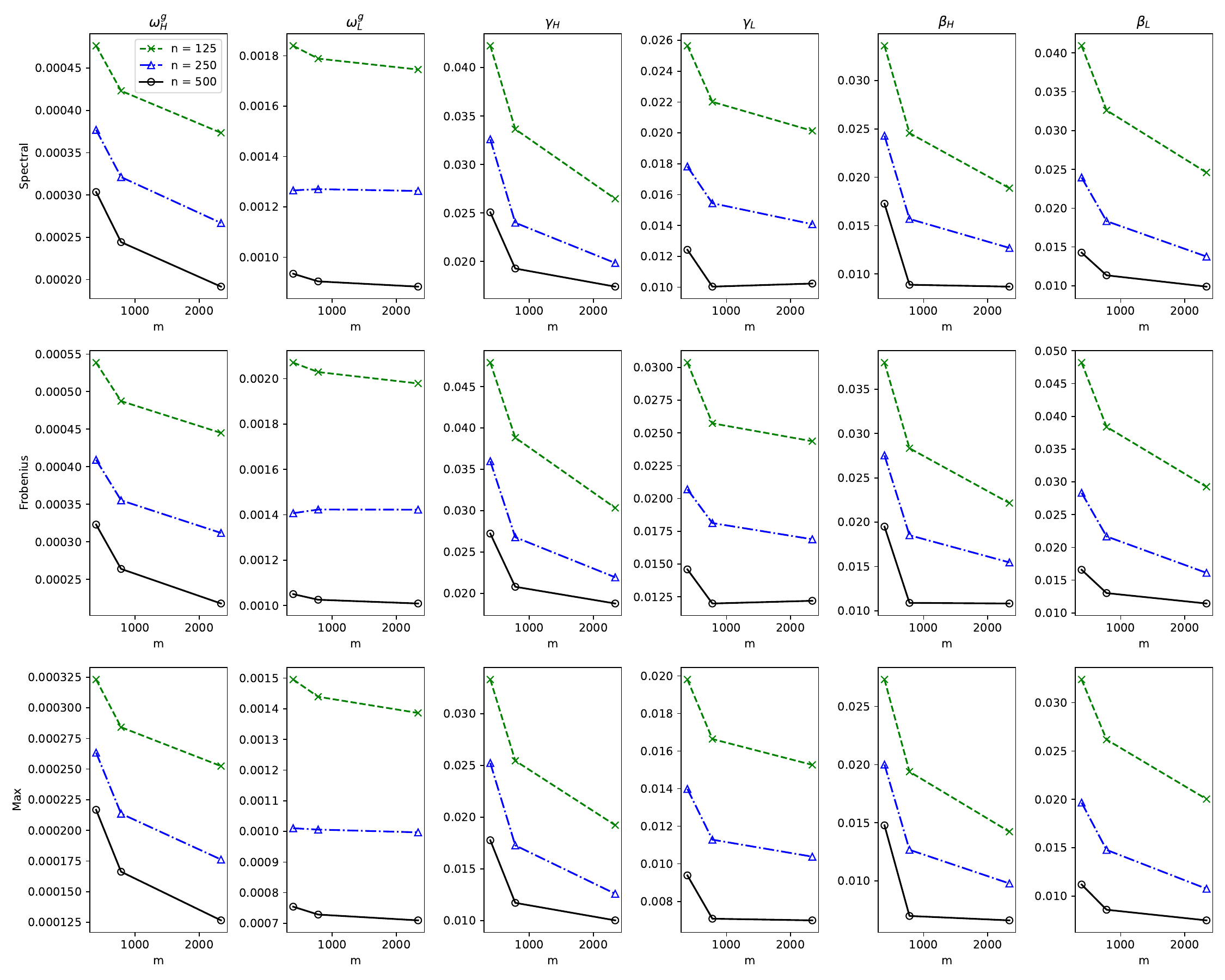}
\caption{Mean matrix spectral norms, Frobenius norms, and max norms of $\hat{\bomega}^g_H - \bomega^g_{H0}$, $\hat{\bomega}^g_L-\bomega^g_{L0}$, $\hat{\bgamma}_H-\bgamma_{H0}$, $\hat{\bgamma}_L-\bgamma_{L0}$, $\hat{\bbeta}_H-\bbeta_{H0}$, and $\hat{\bbeta}_L-\bbeta_{L0}$ for the proposed FOGI estimator with $n=125,250,500$ and $m=390,780,2340$.
\label{Figure:Parameter Large}}
\end{figure}

Figure \ref{Figure:Parameter Large} depicts the mean matrix spectral norms, Frobenius norms, and max norms of $\hat{\bomega}^g_H - \bomega^g_{H0}$, $\hat{\bomega}^g_L-\bomega^g_{L0}$, $\hat{\bgamma}_H-\bgamma_{H0}$, $\hat{\bgamma}_L-\bgamma_{L0}$, $\hat{\bbeta}_H-\bbeta_{H0}$, and $\hat{\bbeta}_L-\bbeta_{L0}$ for the proposed FOGI estimator with $n=125,250,500$ and $m=390,780,2340$.
From Figure \ref{Figure:Parameter Large}, we find that the mean matrix estimation errors decrease as the number of high-frequency observations or low-frequency observations increases, except for the $\hat{\bomega}^g_L - \bomega^g_{L0}$ and $\hat{\bgamma}_L - \bgamma_{L0}$.
We note that there is an unclear effect of increasing the number of high-frequency observations in the case of $\bomega_{L}$ and $\bgamma_{L}$.
This may be because the parameters $\bomega_{L}$ and $\bgamma_{L}$ affect the close-to-open periods, which provide the low-frequency observations, so high-frequency observations have relatively little effect on the estimation accuracy.

We as well studied the prediction performance of the proposed model and compared it with benchmarks.
Specifically, we estimated the conditional large volatility matrix $\E(\bGamma_{n+1} | \mathcal{F}_n)$ by $\tilde{\bGamma}_{n+1}$ and computed the matrix estimation errors for $\tilde{\bGamma}_{n+1} - \E(\bGamma_{n+1} | \mathcal{F}_n)$ in spectral, Frobenius, max, and relative Frobenius norm. 
To construct $\tilde{\bGamma}_{n+1}$, we need to obtain the idiosyncratic volatility matrix estimator $\hat{\bGamma}^s$, where we took $\sqrt{\log p/n}  + \sqrt{1/p}$ to be the threshold. 
For comparisons, we considered the POET procedure proposed by \cite{fan2013large} and the factor GARCH-It\^o (FGI) procedure, which is the same as that of the FOGI procedure based on the factor It\^{o} diffusion process, except for the factor estimation that utilizes the scaled FGI estimator.
For example, the FGI model only considers open-to-close volatility.
If we set parameters in the FOGI model as $\tau = 1$, $\gamma_L = \bI_r$, and $\beta_L = \bzero$, then the FOGI model becomes the FGI model.
Under the FGI model, the conditional open-to-close volatility can be estimated as follows:
\begin{equation*}
	\hat{\bh}_{n+1}(\hbtheta) = \vec (\bomega) + \bR^g \hat{\bh}_{n}(\hbtheta) + \bA^g \vec (RV_{n}),
\end{equation*}
where $\hbtheta = (\vech(\bomega), \vech(\bgamma), \vech(\bbeta))$, $\bR = \bgamma \otimes \bgamma$, $\bB = \bbeta \otimes \bbeta$, $\bvarrho_1 = \bB^{-1} (e^{\bB} - \bI_{r^2})$, $\bvarrho_{2}=\bB^{-2} (e^{\bB} -\bI_{r^2}-\bB ) $, $\bvarrho_{3} =\bB^{-3} (e^{\bB} -\bI_{r^2}-\bB-\bB^2/2)$, $\bvarrho=2\bvarrho_{3}    \bR   +    \bvarrho_{1} - \bvarrho_{2} $, $\bR^g = \bvarrho \bR \bvarrho^{-1}$, and $\bA^g = \bvarrho \bB$.
The model parameter can be estimated by the least squares estimators as follows:
\begin{equation*}
	\hbtheta = \arg \min_{\btheta} \frac{1}{2n} \sum_{k=1}^{n} \{ \vech(RV_k) - \hat{\bh}_k (\btheta) \}^\top \{ \vech(RV_k) - \hat{\bh}_k (\btheta) \}.
\end{equation*}
To match the estimated open-to-close volatility with the daily integrated volatility, one can scale up the estimated open-to-close volatility as $\Lambda \hat{\bh}_{n+1} \Lambda$, where $\Lambda$ is a diagonal matrix such that $\Lambda = \diag  (\sqrt{ ( \[\overline{RV}_{n}\]_{11} + \[\overline{OV}_{n}\]_{11})  \[\overline{RV}_{n}\]_{11} ^{-1}}   ,\ldots, \sqrt{ ( \[\overline{RV}_{n}\]_{pp} + \[\overline{OV}_{n}\]_{pp} )  \[\overline{RV}_{n}\]_{pp} ^{-1} }  )$, $\overline{RV}_{n}$ $  = \frac{1}{n} \sum_{k=1}^{n} RV_{k}$, $\overline{OV}_{n} = \frac{1}{n} \sum_{k=1}^{n} OV_{k}$, and $OV_{k} = \hat{\br}_k \hat{\br}_k^\top$.
We utilized $\hat{\bGamma}_n^H + [ \bX(n) - \bX(n-1) ] [ \bX(n) - \bX(n-1) ]^\top$ for the open-to-open covariance matrix in the POET procedure, and we took $\sqrt{2 \log p/ m^{1/2}}$ to be the threshold for the thresholding step in the POET procedure.

\begin{figure}[t!]
	\centering
	\includegraphics[width=0.85\textwidth]{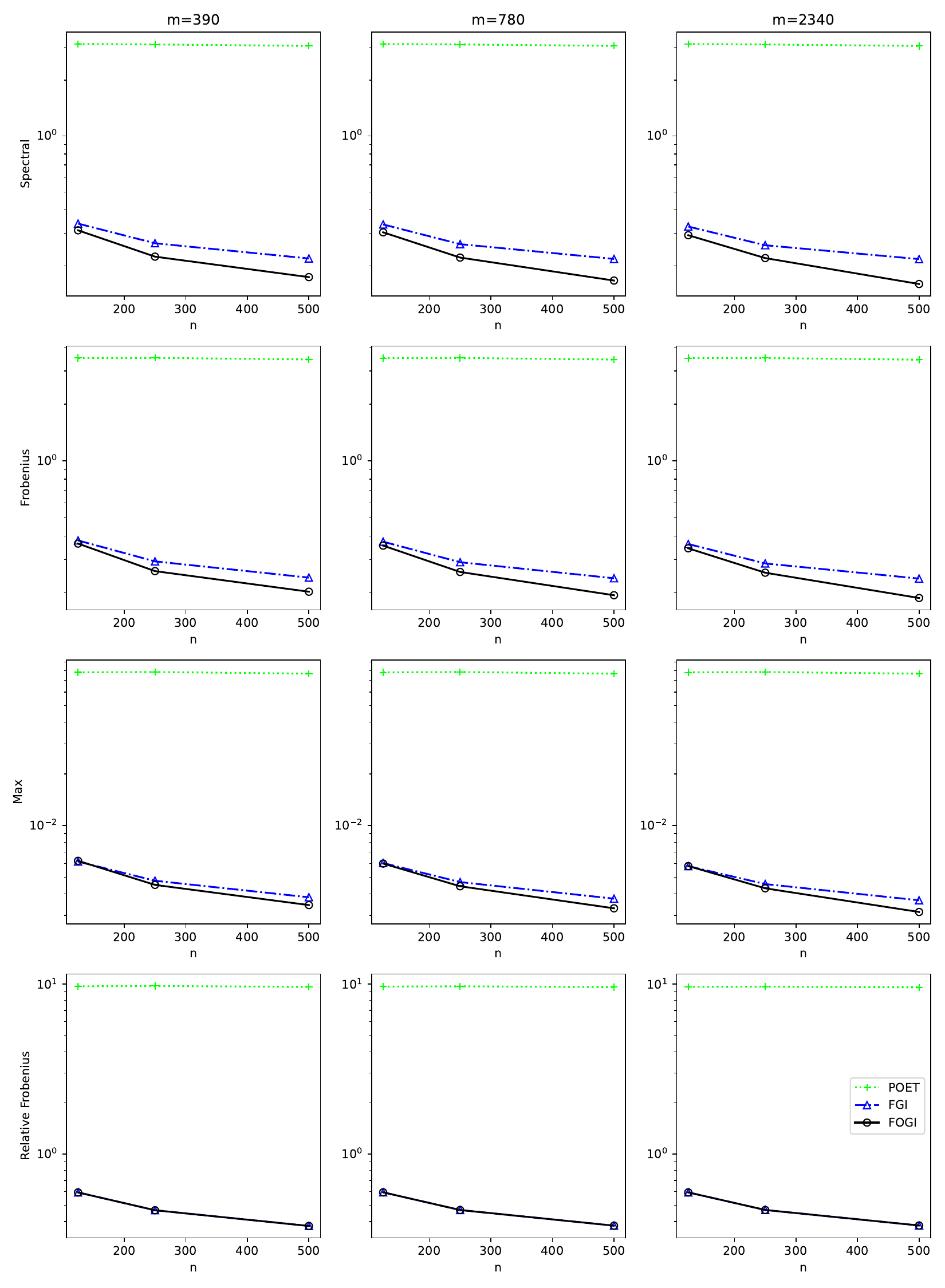}
	\caption{Mean matrix prediction errors in the spectral, Frobenius, max, and relative Frobenius norms for the conditional daily integrated volatility matrix $\E \( \bGamma_{n+1} \middle | \FF_{n}  \)$ with the FOGI, FGI, and POET estimators with $n=125,250,500$ and $m=390,780,2340$. 
	\label{Figure:Forecast Large}}
\end{figure}

Figure \ref{Figure:Forecast Large} draws the mean matrix prediction errors in the spectral norms, Frobenius norms, max norms, and relative Frobenius norms with the FOGI, FGI, and POET estimators with $n=125,250,500$ and $m=390,780,2340$. 
It is clear that the estimator based on the proposed FOGI method outperforms the benchmarks based on the POET and FGI procedures, except for the relative Frobenius norm.
This may be because each eigenvalue in the relative Frobenius norm has a similar scale, and so the effect of the factor part is relatively small. 
Thus, FOGI cannot present significant benefits in terms of  the relative Frobenius norm.
For small $n$, the FGI procedure shows better performance than the FOGI procedure in terms of the max norm.
One of the possible explanations is that since FOGI is more complicated than FGI, for small $n$, FOGI may have larger estimation errors than FGI.
That is, for small $n$, the estimation error is larger than the model error.   
As the number of low- or high-frequency observations increases, the mean matrix prediction errors decrease, which supports the theoretical results derived in Section \ref{SEC-Large}.

\section{Empirical analysis} 
\label{SEC-Empirical}

In this section, we apply the proposed FOGI model to real trading stock prices to examine the case for a large number of assets with $p=200$ and focus on constrained portfolio allocation problems that are widely considered in financial applications.
The data set consists of the minute-by-minute stock prices of companies that were traded on the New York Stock Exchange during the 754 trading days from January 2017 to December 2019. 
That is, in order to synchronize the observation time points, we used the previous tick \citep{zhang2011estimating} scheme.
The high-frequency trading data for the top 200 large volume stocks among the S\&P 500 compositions was obtained from the Wharton Data Service (WRDS) system.
The 6.5 trading hours daily result in $\tau=6.5/24$.

To forecast the one-day-ahead conditional volatility matrix, we employed the FOGI, FGI, and POET procedures explained in Section \ref{SEC-Simulation}, where the only difference is a threshold level for the idiosyncratic volatility matrix estimators.
Here, we used the global industry classification standard (GICS) for sectors and kept the volatilities within the same sector, but set others to be zero \citep{fan2016incorporating}. 
We set the in-sample period as 126 days and used the rolling window scheme to estimate the one-day-ahead conditional volatility matrix for the FOGI and FGI models.

For the empirical case, the decomposition of daily variation into its continuous and jump components can help to explain the volatility dynamics due to the existence of price jumps \citep{ait2012testing,andersen2007roughing,barndorff2006econometrics,corsi2010threshold}.
Therefore, we employed the jump truncated PRVM estimator proposed by \citet{ait2016increased} to obtain $\hat{\bGamma}^H_k$, $k=1,\ldots,n$.
Specifically, jump truncated PRVM were estimated as follows:
\begin{equation*}
	\[\hat{\bGamma}^H_k\]_{ij} = \frac{1}{w \phi} \sum_{u=1}^{m - w + 1} \left (\bar{X}_{k,u}^{i} \bar{X}_{k,u}^{j} - \frac{1}{2} \hat{X}_{k,u}^{ij}  \right ) \mathbbm{1}\left(\abs{\bar{X}_{k,u}^{i} < v_k^{i} }\right) \mathbbm{1}\left(\abs{\bar{X}_{k,u}^{j} < v_k^{j} }\right),
\end{equation*}
where $w = [m^{1/2}]$, $\phi = 1/12$, $g(x)=x \land (1-x)$, $\Delta^{k}_{u}X_{i} = X_{i}(k-1+\tau u/m) - X_{i}(k-1+\tau (u-1)/m)$, 
$\hat{X}_{k,u}^{ij}= \sum_{s=1}^{w} (g(\frac{s}{w}) - g(\frac{s-1}{w}))^2 \Delta^{k}_{u+s-1} X_{i} \Delta^{k}_{u+s-1} X_{j}$, $\bar{X}_{k,u}^{i}= \sum_{s=1}^{w-1} g(s/w) \Delta^{k}_{u+s} X_{i}$, and
$v_k^{i} = 2.19 \sqrt{\sum_{u=1}^{m-w+1} ( \bar{X}_{k,u}^{i} )^2 / (m-w+1)}$ is a truncation parameter.

\begin{figure}[!t]
	\centering
	\includegraphics[width=\textwidth]{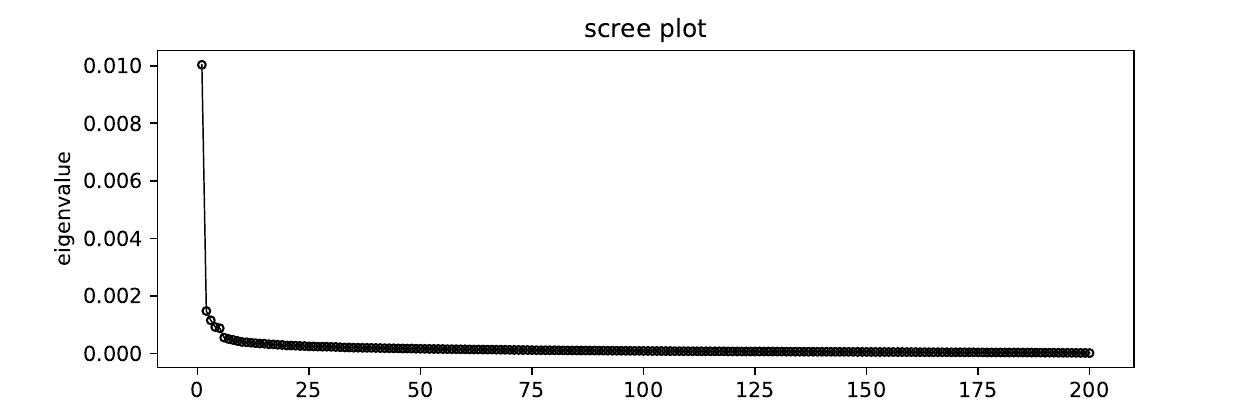}
	\caption{The scree plot of eigenvalues of the average of 754 daily PRVM estimators.
	\label{Figure:Scree Plot}}
\end{figure}

We then need to determine the number of market factors $r$ so that the corresponding factor loading and factor volatility matrices can be identified and estimated.
Figure \ref{Figure:Scree Plot} shows the scree plot for $\hat{\lambda}_1$, $\hat{\lambda}_2$, $\ldots$, $\hat{\lambda}_{200}$, where $\hat{\lambda}_{j}$ is the $j$th largest eigenvalue of $\hat{\bar{\bGamma}}^{H}_n+\hat{\bar{\bGamma}}^{L}_n$ and
\begin{equation*}
	\hat{\bar{\bGamma}}^{H}_n =  \frac{1}{n} \sum_{k=1}^n \hat{\bGamma}_k ^H,  \quad \hat{\bar{\bGamma}}^{L}_n = \frac{1}{n} \sum_{k=1}^{n} (\bX(k) -\bX(\tau+k-1) -\bar{\bX} )  (\bX(k) -\bX(\tau+k-1)  -\bar{\bX} ) ^{\top}
	.
\end{equation*}
From Figure \ref{Figure:Scree Plot}, we find that possible candidates for the number of market factors $r$ is 1,2,3,4. 
To further determine the rank $r$, we adopted the procedure as described in \cite{ait2015using}, 
\begin{equation*}
	\hat{r} = \arg \underset{1 \leq j \leq r_{\max}}{\min} \[ K(\hat{\bar{\bGamma}}^{H}_n, n, m, j) + K(\hat{\bar{\bGamma}}^{L}_n, n, 1, j) \] ,
\end{equation*}
where
\begin{align*}
  &K(\bGamma,n,m,j) = p^{-1}  \lambda_{j} + j \times c_1 \left\lbrace \sqrt{\log p / (nm^{1/2})} + p^{-1} \log p \right\rbrace^{c_2}  -1
\end{align*}
and $\lambda_{j}$ is the $j$th largest eigenvalue of $\bGamma$.
We studied the problem when $r_{\max}=30$, $c_1=0.02 \hlambda_{k,30}$, and $c_2=0.5$. 
The procedure chose $\hat{r}=3$.

We first investigate the effects of the overnight risk.
We can find that the magnitude of overnight factor volatility is comparable to that of intraday factor volatility by calculating the intraday factor volatility ratio of the total factor volatility as follows:
\begin{equation*}
  \frac{\lVert{\bar{IV}}\rVert_{2}}{\lVert{\bar{IV}}\rVert_{2}+\lVert{\bar{OV}}\rVert_{2}} = 0.501
  ,
\end{equation*}
where $\bar{IV}$ and $\bar{OV}$ are the average of estimated intraday and overnight factor volatilities, respectively.
Since, in terms of magnitude, the overnight factor volatility plays a significant role as much as the intraday factor volatility, using the overnight information might be helpful to capture the whole-day dynamics.
To check this, we compared the estimated parameters for the intraday and overnight risks, $\bB^{g}/\tau$ and $\bA^{g}/\tau$.
\begin{figure}[!h]
\centering
\includegraphics[width = 0.9\textwidth]{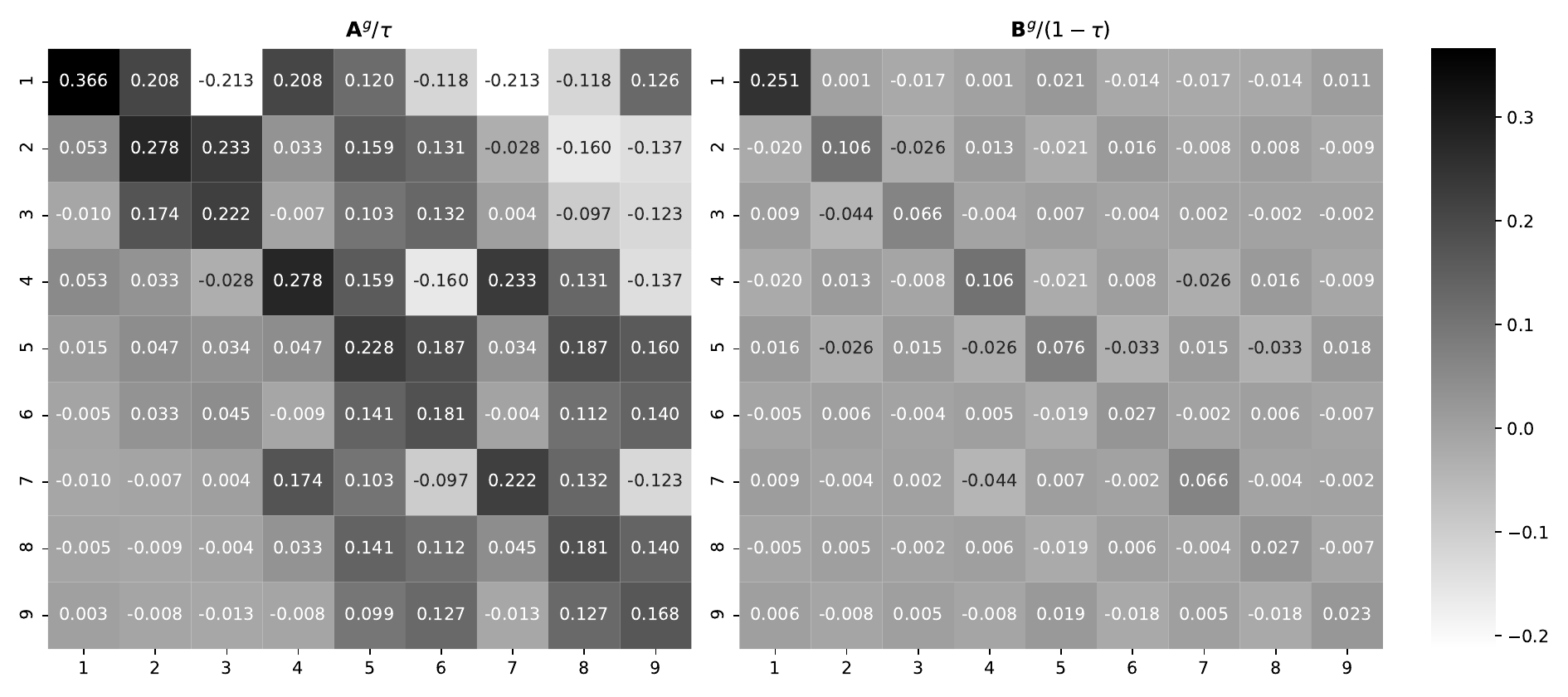}
\caption{Estimated parameters of the FOGI model using the whole data points.} \label{fig:estimated}
\end{figure}
Figure \ref{fig:estimated} draws the estimated parameters of the FOGI model using the whole data points.
From Figure \ref{fig:estimated}, we can find that the first factor--considered the market factor--of overnight volatility helps capture the whole-day factor volatility dynamics, whereas all three factors of intraday volatility significantly affect the future whole-day factor volatility dynamics.
Since the largest factor of overnight volatility accounts for a significant portion of whole-day factor volatility, we can conjecture that incorporating overnight information is beneficial.

We revisited the constrained portfolio allocation problem \citep{fan2012vast}.
Given the estimated predictor $\tilde{\bGamma}_{k}$ for a future large volatility matrix, we minimized the following portfolio risk function:
	\begin{equation*}
	\underset{\bw_k \text{ s.t. } \bw_k^{\top} \mathbf{J}=1 \text{ and } \| \bw_k \|_1 \leq c_0 }{\text{min}} \bw_k^{\top} \tilde{\bGamma}_{k} \bw_k,
	\end{equation*}
where $\mathbf{J}=(1, \ldots, 1)^{\top} \in \RR^p$ and $c_0$ is the gross exposure constraint that ranges from 1 to 3. 
The portfolio associated with $\hat{\bw}_k$ that minimizes the above function is the so-called optimal portfolio. 
For a given period with $d$ days, we computed the out-of-sample portfolio risk for the optimal portfolios in the annualized form
\begin{equation*}
	R = \sqrt{\frac{252}{d} \sum_{k=1}^{d} \left( \sum_{i=1}^{39} r_{k,i}(\hat{\bw}_k)^2 + r_{k}^{CO}(\hat{\bw}_k)^2 \right)},
\end{equation*}
where $r_{k,i}(\hat{\bw}_k)  = \hat{\bw}_k^{\top} (\bY(k-1+\tau\frac{i}{39}) - \bY(k-1+\tau\frac{i-1}{39}))$ is the 10-min portfolio log-return and $r_{k}^{CO}(\hat{\bw}_k) = \hat{\bw}_k^{\top} (\bY(k) - \bY(k-1+\tau))$ is the close-to-open portfolio log-return of day $k$.
We used six different out-of-sample periods--day 127 to day 252, day 253 to day 378, day 379 to day 504, day 505 to day 630, and day 630 to day 754--and the whole out-of-sample period.

\begin{figure}[!t]
\centering
\includegraphics[width=\textwidth]{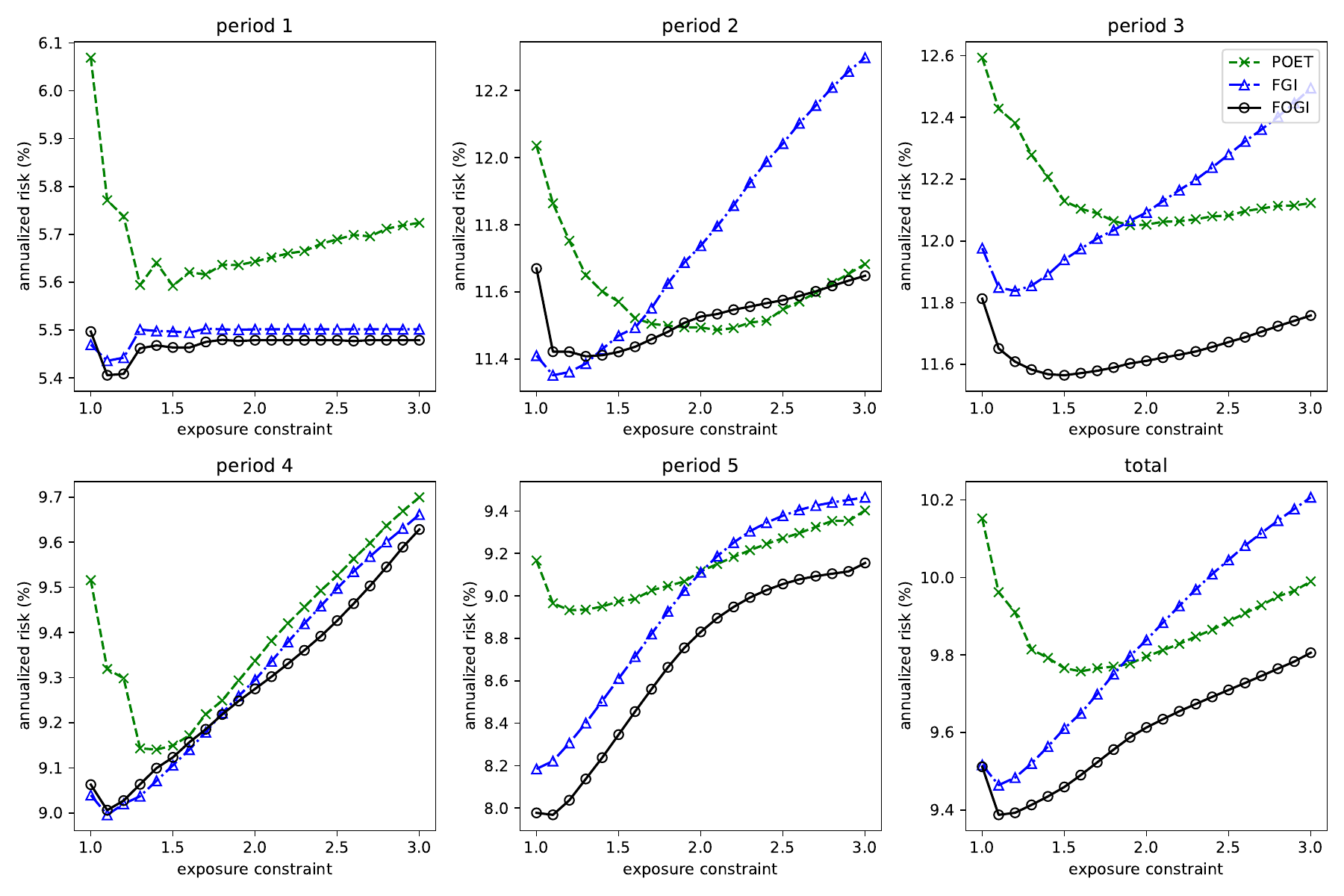}
\caption{Annualized out-of-sample portfolio risks with the FOGI, FGI, and POET procedures against the gross exposure constraint $c_0$ for the five different periods and the whold period.
\label{Figure: empirical-portfolio}}
\end{figure}

Figure \ref{Figure: empirical-portfolio} plots the annualized out-of-sample portfolio risks with the FOGI, FGI, and POET procedures for the six different periods--say, period 1, 2, 3, 4, and 5--and the whole period against the gross exposure constraint $c_0$.  
From Figure \ref{Figure: empirical-portfolio}, we find that the parametric models, the FOGI and FGI models, perform better than the POET method.
It may be because the factor has a time series dynamic structure.
We note that the proposed FOGI model usually has the smallest portfolio risks among the methods except for periods 2 and 4.
For period 4, the difference is not significant, while for period 2, the difference is relatively big.
This may be because during period 2, the G2 trade war occurred; thus, there is a structural break \citep{oh2023effect}. 
This makes it hard to estimate parameters; thus, the relatively complicated FOGI model may show worse performance in this period. 
For period 4, the overnight information may be relatively less important. 
For the whole period, FOGI shows the best performance. 
From these results, we may conjecture that the market factors have a time series dynamic structure and that incorporating close-to-open market factors helps to account for market dynamics.



To check if the above results are robust to the different choices of rank, we conducted the same constrained portfolio allocation problem with the different choices of rank.
\begin{figure}[!h]
\centering
\includegraphics[width = 1\textwidth]{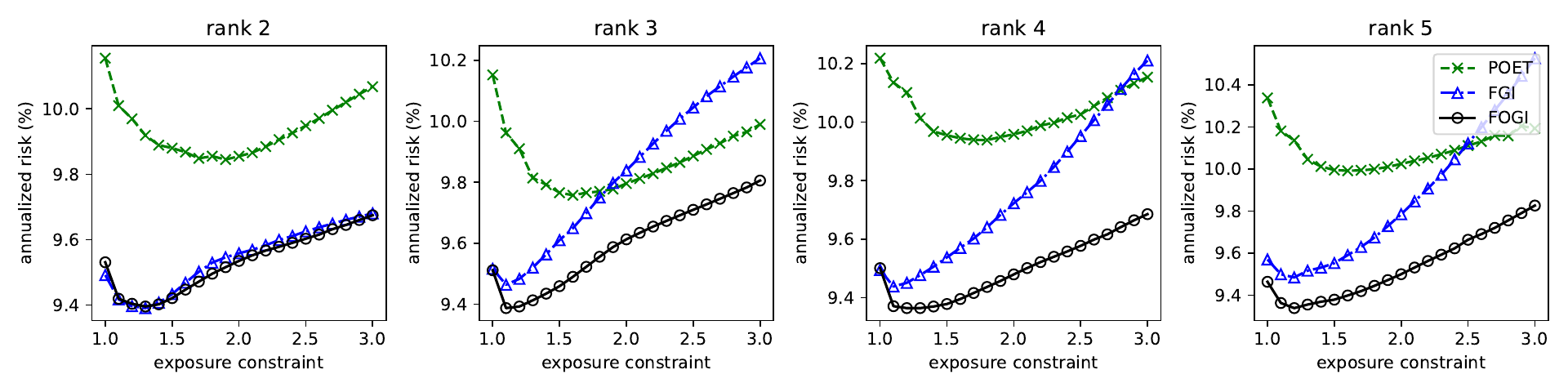}
\caption{Annualized out-of-sample portfolio risks with the FOGI, FGI, and POET procedures against the gross exposure constraint $c_0$ for the different choice of the rank.}
\label{fig:rank}
\end{figure}
Figure \ref{fig:rank} plots the annualized out-of-sample portfolio risks with the FOGI, FGI, and POET procedures for the different choices of the rank against the gross exposure constraint $c_0$.
Figure \ref{fig:rank} shows that the FOGI and FGI models perform similarly regardless of the chosen rank among proper candidates.
FGI shows the best performance for the rank 2 and the performance is relatively unstable for the choice of the rank. 
In contrast, FOGI shows relatively stable results regardless of the choice of rank.
Therefore, we can conclude that the results in Figure 4 are robust to the different choices of the number of factors.



\section{Conclusion} \label{SEC-Conclusion}

In this paper, we propose the unified approach for describing the volatility matrix evolution by embedding the discrete-time multivariate GARCH model structure in the continuous-time It\^o diffusion process. 
We model the stock price process by the factor It\^o diffusion process so that the large volatility process can be decomposed into the factor and idiosyncratic volatility matrices. 
We embed the multivariate GARCH model structure in the instantaneous factor volatility process, and we assume that the idiosyncratic volatility process is constant over time.
Model parameters in the FOGI model are estimated by minimizing the weighted loss function, and this estimating approach is proved to have good asymptotic behaviors. 
We demonstrate the validity of the proposed estimation method via the simulation study, and the empirical study presents the advantages of the proposed FOGI model in portfolio allocation problems. 
 Thus, this study provides the intuition for the large volatility matrix dynamics of the  open-to-close and close-to-open volatilities.

 \section*{Funding}

  The research of Donggyu Kim was supported in part by the National Research Foundation of Korea (NRF)
[2021R1C1C1003216].
The research of Yazhen Wang was supported in part by NSF grant DMS-1913149.

\bibliography{myReferences}
\end{doublespace}

\end{document}


\maketitle

\begin{spacing}{1.45}

\appendix

    \section{Proofs}
    \label{SEC-Proof}
    
    \subsection{Proof of Theorem \ref{Prop-IV}}
    
    \textbf{Proof of Theorem \ref{Prop-IV}.}
    
    First consider $(a)$. 
    It\^o's lemma provides
        \begin{eqnarray*}  
        \bG_H  (k)  &\equiv& \int _{n-1} ^{\tau+ n-1} \frac {(\tau+n-1-t) ^k}{k!}  \vec (\bSigma_t  ) dt \cr
        &=&  \frac{2 \tau^{k+1}}{(k+3)!} \bR_H \vec(\bomega_{H1}) -   \frac{\tau^{k+1}}{(k+2)!} \vec(\bomega_{H2}) \cr
        && +  \left \{ \frac{2 \tau^{k+1}}{(k+3)!}  \bR_H   + \( \frac{\tau^{k+1}}{(k+1)!} -    \frac{\tau^{k+1}}{(k+2)!} \)  \bI_{r^2}  \right \}\vec (\bSigma_{n-1} )  \cr
        &&+\tau \(  \frac{\tau^{k}}{(k+2)! } - 2  \frac{\tau^{k}}{(k+3)!} \)  \vec ( \bnu \bnu ^{\top})\cr 
        &&+ \frac{ \bnu \otimes \bnu }{ \tau^{ 2}} \vec \(  \(A_{k,ij} + A_{k, ji}  \)_{i,j=1,\ldots, r} \)    + \tau^{-1} \bB_H    \bG_H (k+1) \text{ a.s.},
        \end{eqnarray*}
    where $A_{k,ij} =\int _{n-1} ^{\tau +n-1} \frac{ (\tau+n-1-t)^{k+2} }{(k+2)\times k!} Z_{i,t}  dZ_{j,t} $.
    Then we have
        \begin{eqnarray*}  
        \bG_H  (0)  &=& \int _{n-1} ^{\tau+ n-1}   \vec (\bSigma_t  ) dt \cr
        &=& \sum_{k=0}^{\infty}   \frac{2 \tau}{(k+3)!} \bB_H ^{k} \bR_H \vec(\bomega_{H1}) -   \frac{\tau}{(k+2)!} \bB_H ^{k} \vec(\bomega_{H2})  \cr
        &&+ \sum_{k=0}^{\infty}  \tau  \left \{ \frac{2  }{(k+3)!}  \bB_H^k \bR_H   + \( \frac{1}{(k+1)!} -   \frac{ 1}{(k+2)!} \)  \bB_H^k  \right \} \vec (\bSigma_{n-1} )  \cr
        &&+ \sum_{k=0}^{\infty} \tau  \( \frac{1}{(k+2)! } - 2  \frac{1}{(k+3)!} \) \bB_H^k  \vec ( \bnu \bnu ^{\top})\cr 
        &&+  \sum_{k=0}^{\infty} \frac{ \bnu \otimes \bnu }{ \tau^{ 2}} \bB_H ^k\vec \(  \(A_{k,ij} + A_{k, ji}  \)_{i,j=1,\ldots, r}     \)   \cr
        &=&  \tau \big[  2  \bvarrho_{H3} \bR_H  \vec(\bomega_{H1}) -  \bvarrho_{H2}   \vec(\bomega_{H2})   + \(\bvarrho _{H2} - 2 \bvarrho_{H3}  \) \vec ( \bnu \bnu ^{\top}) \cr
        &&+ \(2\bvarrho_{H3}    \bR_H   +    \bvarrho_{H1} - \bvarrho_{H2}    \)\vec (\bSigma_{n-1} )  \big ] \cr
        &&+  \frac{ \bnu \otimes \bnu }{ \tau^{ 2}}  \sum_{k=0}^{\infty}\bB_H ^k\vec \(  \(A_{k,ij} + A_{k, ji}  \)_{i,j=1,\ldots, r}     \) \text{ a.s.}
        \end{eqnarray*}
    Then we have
        \begin{eqnarray*}
        &&\bvarrho_H ^{-1} \bh_n^H (\btheta)  \cr
        &&= 	\bvarrho_H ^{-1}  \[ 2  \bvarrho_{H3} \bR_H  \vec(\bomega_{H1}) -  \bvarrho_{H2}   \vec(\bomega_{H2})   + \(\bvarrho _{H2} - 2 \bvarrho_{H3}  \) \vec ( \bnu \bnu ^{\top}) \]  +   \vec (\bSigma_{n-1}) \cr
        &&=\bvarrho_H ^{-1}  \[ 2  \bvarrho_{H3} \bR_H  \vec(\bomega_{H1}) -  \bvarrho_{H2}   \vec(\bomega_{H2})   + \(\bvarrho _{H2} - 2 \bvarrho_{H3}  \) \vec ( \bnu \bnu ^{\top}) \]   \cr
        && \quad+ \vec \(  \bomega_L \) + \bR_L \vec \(\bomega_H \) +  \frac{1}{\tau} \bR_L \bB_H \int_{n-2}^{\tau +n-2} \vec \( \bSigma_t \) dt  + \frac{1}{1-\tau} \bB_L  \vec \( \br _{n-1} \br_{n-1}^{\top} \) \cr
        &&\quad +  \bR_L \bR_H  \vec \( \bSigma _{n-2}\)   \cr
        && =\(\bI_{r^2} - \bR_L \bR_H \)  \bvarrho_H ^{-1}  \[ 2  \bvarrho_{H3} \bR_H  \vec(\bomega_{H1}) -  \bvarrho_{H2}   \vec(\bomega_{H2})   + \(\bvarrho _{H2} - 2 \bvarrho_{H3}  \) \vec ( \bnu \bnu ^{\top}) \]  \cr
        && \quad +  \vec \(  \bomega_L \) + \bR_L \vec \(\bomega_H \) \cr
        &&\quad + \bR_L \bR_H \bvarrho_H ^{-1} \bh_{n-1} ^H (\btheta)  +\tau^{-1} \bR_L \bB_H \int_{n-2}^{\tau +n-2} \vec \( \bSigma_t \) dt  + (1-\tau)^{-1} \bB_L  \vec \( \br _{n-1} \br_{n-1}^{\top} \).
        \end{eqnarray*}
    Thus, we have
        \begin{eqnarray}
        \label{Thm1-eq1}
        && \bh_n^H (\btheta)  \cr
        && =\bvarrho_H \(\bI_{r^2} - \bR_L \bR_H \)  \bvarrho_H ^{-1}  \[ 2  \bvarrho_{H3} \bR_H  \vec(\bomega_{H1}) -  \bvarrho_{H2}   \vec(\bomega_{H2})   + \(\bvarrho _{H2} - 2 \bvarrho_{H3}  \) \vec ( \bnu \bnu ^{\top}) \]  \cr
        && \quad +  \bvarrho_H  \vec \(  \bomega_L \) + \bvarrho_H \bR_L \vec \(\bomega_H \) \cr
        &&\quad +\bvarrho_H  \bR_L \bR_H \bvarrho_H ^{-1} \bh_{n-1} ^H (\btheta)  +\frac{1}{\tau} \bvarrho_H  \bR_L \bB_H \int_{n-2}^{\tau +n-2} \vec \( \bSigma_t \) dt  + \frac{1}{1-\tau}\bvarrho_H  \bB_L  \vec \( \br _{n-1} \br_{n-1}^{\top} \) \cr
        &&= \vec \( \bomega_{H}^g\) + \bR_H^g  \bh_{n-1} ^H (\btheta)  +\frac{  \bA_H^g }{\tau} \int_{n-2}^{\tau +n-2} \vec \( \bSigma_t \) dt  +  \frac{\bB_H^g}{1-\tau} \vec \( \br _{n-1} \br_{n-1}^{\top} \).
        \end{eqnarray}
    Similarly we can show $(b)$.
    Consider $(c)$.
    By $(a)$ and $(b)$, we have
        \begin{eqnarray*}
        &&\int_{n-1}^{n} \vec \( \bSigma_t  \) dt \cr
        &&=   \tau\[ 2  \bvarrho_{H3} \bR_H  \vec(\bomega_{H1}) -  \bvarrho_{H2}   \vec(\bomega_{H2})   + \(\bvarrho _{H2} - 2 \bvarrho_{H3}  \) \vec ( \bnu \bnu ^{\top}) \]  \cr
        && \quad + (1-\tau) \[ 2  \bvarrho_{L3} \bR_L  \vec(\bomega_{L1}) -  \bvarrho_{L2}   \vec(\bomega_{L2})   \]   \cr
        &&\quad +\tau \bvarrho _H \vec (\bSigma_{n-1} ) + (1-\tau)  \bvarrho_L  \vec (\bSigma_{\tau+n-1}) +\bQ_n ^H + \bQ_n ^L \cr
        &&=  \tau\[ 2  \bvarrho_{H3} \bR_H  \vec(\bomega_{H1}) -  \bvarrho_{H2}   \vec(\bomega_{H2})   + \(\bvarrho _{H2} - 2 \bvarrho_{H3}  \) \vec ( \bnu \bnu ^{\top}) \]  \cr
        && \quad + (1-\tau) \[ 2  \bvarrho_{L3} \bR_L  \vec(\bomega_{L1}) -  \bvarrho_{L2}   \vec(\bomega_{L2})   \] +  (1-\tau )  \bvarrho_L  \vec \( \bgamma_H  \bomega_{H1} \bgamma_H ^{\top} -  \bomega_{H2}   \)  \cr
        &&\quad  + (\tau \bvarrho _H  +   (1-\tau) \bvarrho_L\bR_H ) \vec (\bSigma_{n-1} ) + \frac{1-\tau}{ \tau }  \bvarrho_L \bB_H \int_{n-1}^{\tau+n-1} \vec \( \bSigma_t\) dt  \cr
        &&\quad  +\bQ_n ^H + \bQ_n ^L \cr 
        &&=    \tau\[ 2  \bvarrho_{H3} \bR_H  \vec(\bomega_{H1}) -  \bvarrho_{H2}   \vec(\bomega_{H2})   + \(\bvarrho _{H2} - 2 \bvarrho_{H3}  \) \vec ( \bnu \bnu ^{\top}) \]  \cr
        && \quad + (1-\tau) \[ 2  \bvarrho_{L3} \bR_L  \vec(\bomega_{L1}) -  \bvarrho_{L2}   \vec(\bomega_{L2})   \] +  (1-\tau )  \bvarrho_L  \vec \( \bgamma_H  \bomega_{H1} \bgamma_H ^{\top} -  \bomega_{H2}   \)  \cr
        && \quad + (\tau \bvarrho _H  +   (1-\tau) \bvarrho_L\bR_H ) \vec (\bSigma_{n-1} ) +( 1-\tau)  \bvarrho_L \bB_H \bh_n^H (\theta)  \cr
        && \quad + \(  (1-\tau)  \tau ^{-1}  \bvarrho_L \bB_H  + \bI_{r^2} \)\bQ_n ^H + \bQ_n ^L \cr 
        &&=  \left \{ ( 1-\tau)  \bvarrho_L \bB_H +\tau \bI_{r^2} \right \} \[ 2  \bvarrho_{H3} \bR_H  \vec(\bomega_{H1}) -  \bvarrho_{H2}   \vec(\bomega_{H2})   + \(\bvarrho _{H2} - 2 \bvarrho_{H3}  \) \vec ( \bnu \bnu ^{\top}) \] \cr
        &&\quad + (1-\tau) \[ 2  \bvarrho_{L3} \bR_L  \vec(\bomega_{L1}) -  \bvarrho_{L2}   \vec(\bomega_{L2})   \] +  (1-\tau )  \bvarrho_L  \vec \( \bgamma_H  \bomega_{H1} \bgamma_H ^{\top} -  \bomega_{H2}   \)  \cr
        &&\quad  + \{ \tau \bvarrho _H  +   (1-\tau) \bvarrho_L\bR_H + (1-\tau) \bvarrho_L \bB_H \bvarrho_H \} \vec (\bSigma_{n-1} )  + \bQ_n \text{  a.s.}
        \end{eqnarray*}
    Algebraic manipulations show
        \begin{eqnarray*}
        &&\bh_n(\btheta) \cr
        &&=   \left \{ ( 1-\tau)  \bvarrho_L \bB_H +\tau \bI_{r^2} \right \} \[ 2  \bvarrho_{H3} \bR_H  \vec(\bomega_{H1}) -  \bvarrho_{H2}   \vec(\bomega_{H2})   + \(\bvarrho _{H2} - 2 \bvarrho_{H3}  \) \vec ( \bnu \bnu ^{\top}) \] \cr
        &&\quad + (1-\tau) \[ 2  \bvarrho_{L3} \bR_L  \vec(\bomega_{L1}) -  \bvarrho_{L2}   \vec(\bomega_{L2})   \] +  (1-\tau )  \bvarrho_L  \vec \(  \bomega_{H}    \) + \brho \vec (\bSigma_{n-1} )  \cr
        &&=  \vec \( \bomega ^g\) + \bR ^g  \bh_{n-1} ^H (\btheta)  +\frac{  \bA ^g }{\tau} \int_{n-2}^{\tau +n-2} \vec \( \bSigma_t \) dt  +  \frac{\bB ^g}{1-\tau} \vec \( \br _{n-1} \br_{n-1}^{\top} \),
        \end{eqnarray*}
    where the last equality can be derived similar to \eqref{Thm1-eq1}.
    \endpf
    
    \subsection{Proof of Theorem \ref{Thm:Theta-Large}}
    
    Define terminologies: 
        \begin{eqnarray}
        &\hat{L}_{n,m} ( \btheta)& = \frac{1} {2n}\sum _{k=1} ^{n} \Big [   \left \{ \vech(RV_k) -  \tau \hat{\bh}_{h,k}^H (\btheta ) \right \} ^{\top} \hat{\bV}_H ^{-1}  \left \{ \vech(RV_k) -  \tau \hat{\bh}_{h,k}^H (\btheta ) \right \}   \cr
        && \qquad\qquad   + \left \{ \hat{\br}_{h,k}^2 (\bmu_{L}) - (1- \tau) \hat{\bh}_{h,k}^L (\btheta ) \right \} ^{\top} \hat{\bV}_L ^{-1}  \left \{\hat{\br}_{h,k}^2 (\bmu_{L}) -  (1-\tau) \hat{ \bh}_{h,k}^L (\btheta ) \right \}  \Big ] \cr
        &&  = -\frac{1} {2n}\sum _{k=1} ^{n} \hat{l} _k (\btheta) ; \cr
        & L_{n} ( \btheta) &=\frac{1} {2n}\sum _{k=1} ^{n} \Big [   \left \{ \vech(IV_k^f ) -  \tau \bh_{h,k}^H (\btheta ) \right \} ^{\top} \hat{\bV}_H ^{-1}  \left \{ \vech(IV_k^f) -  \tau \bh_{h,k}^H (\btheta ) \right \}   \cr
        &&\qquad\qquad +   \left \{ \br_{h,k}^2 (\bmu_{L}) - (1- \tau) \bh_{h,k}^L (\btheta ) \right \} ^{\top} \hat{\bV}_L ^{-1}  \left \{\br_{h,k}^2 (\bmu_{L}) -  (1-\tau) \bh_{h,k}^L (\btheta ) \right \}  \Big ] \cr
        && =-\frac{1} {2n}\sum _{k=1} ^{n} l _k (\btheta) ; \label{L:def} \\
        & L_{0, n} ( \btheta) &=\frac{1} {2n}\sum _{k=1} ^{n} \Big [   \left \{\tau \bh_{h,k}^H (\btheta_0 ) -  \tau \bh_{h,k}^H (\btheta ) \right \} ^{\top} \hat{\bV}_H ^{-1}  \left \{ \tau \bh_{h,k}^H (\btheta_0 ) -  \tau \bh_{h,k}^H (\btheta ) \right \}   \cr
        &&\qquad \qquad + \Tr \( \hat{\bV}_H ^{-1}  E\[    \bQ_{h,k}^{H }  \bQ_{h,k}^{H \top} \middle | \FF_{k-1} \]+ \hat{\bV}_L ^{-1}  E\[   \bQ_{h,k}^{LL }  \bQ_{h,k}^{LL \top} \middle | \FF_{k-1} \] \)    \cr
        &&\qquad \qquad +   (1- \tau)^2 \left \{  \bh_{h,k}^L (\btheta_0 ) -   \bh_{h,k}^L (\btheta ) \right \} ^{\top} \hat{\bV}_L ^{-1}  \left \{  \bh_{h,k}^L (\btheta_0 )-   \bh_{h,k}^L (\btheta ) \right \}    \cr
        &&\qquad \qquad +   (1- \tau)^4  \vech\( \(  \bmu_{L0} -   \bmu_{L} \)  \(  \bmu_{L0} -   \bmu_{L} \) ^{\top}\) ^{\top} \hat{\bV}_L ^{-1}   \vech\( \(  \bmu_{L0} -   \bmu_{L} \)  \(  \bmu_{L0} -   \bmu_{L} \) ^{\top}\)  \Big ] \cr
        && =-\frac{1} {2n}\sum _{k=1} ^{n} l _{0,k} (\btheta) ; \cr
        &  \hat{\psi}_{n,m} ( \btheta)& =\frac{\partial  \hat{L}_{n,m} ( \btheta) }{\partial \btheta} \quad   \text{ and } \quad  \psi_n ( \btheta)=\frac{\partial  L_{n} ( \btheta) }{\partial \btheta},
        \end{eqnarray}
    where $  \bQ_{h,k} ^H= \vech(IV_k^f) -  \tau\bh _{h,k}^H (\btheta_0 ) $ and $ \bQ_{h,k} ^{LL}  = \br_{h,k}^2 (\bmu_{L0}) -  (1-\tau)  \bh _{h,k}^L (\btheta_0 ) $.
    
    \begin{lemma}
    \label{Lemma:L-consistent}
    Under the assumptions in Theorem \ref{Thm:Theta-Large}, we have
        $$ 
        \sup_{\btheta \in \bTheta} \left | \hat{L}_{n,m}  (\btheta) -L_{0,n}   (\btheta) \right |= o_p(1).
        $$
    \end{lemma}
    
    \textbf{Proof of Lemma \ref{Lemma:L-consistent}.}
    We first show that the estimators $RV_k $ and $\hat{\br}_k(\bmu_{0})$ uniformly converge to the corresponding parameters with the order  $\nu_{m,n} + m^{-1/4}+ n^{-1/2}+ (\pi(p) / p )^{1/2}$.
    To ease the notation, we let $k=1$. 
    We  investigate their convergence rates.
    Let
        \begin{eqnarray*}
        A_1&=&  \{ p^{-1} \hat{\bU} ^{\top}  \(\bX(1)- \bX(\tau) \) - (1-\tau)  \bmu_{L} \}   \{ p^{-1} \hat{\bU} ^{\top} \(\bX(1)- \bX(\tau) \) - (1-\tau) \bmu_{L} \}^{\top}\cr
        && -  \{ \bff(1)- \bff(\tau) +(1-\tau) \bmu_{L0} - (1-\tau) \bmu_{L} \}  \{\bff(1)- \bff(\tau) +(1-\tau) \bmu_{L0}- (1-\tau) \bmu_{L} \} ^{\top}.
        \end{eqnarray*}
    Using the similar arguments of the proofs of Lemma 6.1 \citep{kim2019factor}, we can show
        \begin{equation} 
        \label{Thm-large:eq-01}
        \E  \( \max_{\bmu_{0}} \| A_1 \|_2^b \) \leq C \{ \nu_{m,n} ^b+ (\pi(p) / p )^{b/2} \} .
        \end{equation}
    We consider $p^{-1} \hat{\bU} ^{\top} \hat{\bGamma}_k \hat{\bU} $. 
    Simple algebraic manipulations show
        \begin{eqnarray*}
        \|p^{-1} \hat{\bU}^{\top} \hat{\bGamma}_k \hat{\bU} -\bPsi_k  \|_2 
        &\leq& p^{-2} \|  \hat{\bU} ^{\top} \hat{\bGamma}_k \hat{\bU} -  \bU ^{\top} \bGamma_k \bU  \|_2 + \|p^{-2}  \bU ^{\top} \bGamma_k \bU   -\bPsi_k  \|_2 \cr
        &\leq& p^{-2} \|  \hat{\bU} ^{\top} \hat{\bGamma}_k \hat{\bU} -  \bU ^{\top} \bGamma_k \bU  \|_2+ p^{-2} \| \bU ^{\top} \bGamma_k^s \bU  \|_2.
        \end{eqnarray*}
    For $\|  \hat{\bU}^{\top} \hat{\bGamma}_k \hat{\bU} -  \bU ^{\top} \bGamma_k \bU  \|_2$, we have
        \begin{eqnarray}
        \label{Thm-large:eq-1}
        \|  \hat{\bU}^{\top} \hat{\bGamma}_k \hat{\bU} -  \bU ^{\top} \bGamma_k \bU  \|_2 
        &\leq& \|  \hat{\bU} \|_2^2  \|\hat{\bGamma}_k- \bGamma_k\|_2   + \|\hat{\bU} ^{\top}  \bGamma_k \hat{\bU} - \bU ^{\top}  \bGamma_k \bU  \|_2 \cr
        &\leq& \|  \hat{\bU} \|_2^2  \|\hat{\bGamma}_k- \bGamma_k\|_2   + \|\hat{\bU} - \bU \| _2 \| \bGamma_k\|_2 \| \hat{\bU}\| _2  +\|\hat{\bU} - \bU \| _2 \| \bGamma_k\|_2 \|  \bU\| _2  \cr
        &\leq& O_p( p^2 m^{-1/4} + p^{2}   \nu_{m,n}), 
        \end{eqnarray}
    where the last line is due to Assumption \ref{Assumption1}.
    For $\| \bU ^{\top} \bGamma_k^s \bU  \|_2$,  we have
        \begin{eqnarray}
        \label{Thm-large:eq-2}
        \|  \bU ^{\top} \bGamma_k^s \bU  \|_2 
        &\leq& \| \bU \|_2^2 \| \bGamma_k^s\| _1 \cr
        &\leq& C p \max_{1 \leq j \leq p} \sum_{i=1}^p  |\Gamma_{k,ij} ^s| \cr
        &\leq& C M p \pi(p),
        \end{eqnarray}
    where the last inequality is due to the sparsity condition \eqref{sparsity}.
    By \eqref{Thm-large:eq-1} and \eqref{Thm-large:eq-2}, we have
        \begin{equation}
        \label{Thm-large:eq-03}
        \|p^{-1} \hat{\bU}^{\top} \hat{\bGamma}_k \hat{\bU} -\bPsi_k  \|_2 = O_p ( m^{-1/4}+\nu_{m,n}+ \pi(p)/p ).
        \end{equation}
By \eqref{Thm-large:eq-01} and \eqref{Thm-large:eq-03}, we can show
        \begin{equation*}
        \sup_{\btheta \in \bTheta} \left | \hat{L}_{n,m}  (\btheta) -L_{n}   (\btheta) \right |= o_p(1).
        \end{equation*}
 
    Now it is enough to show $\sup_{\btheta \in \bTheta} \left |  L _{n}  (\btheta) -L_{0, n}   (\btheta) \right |= O_p(n^{-1/2})$.
    We have
        \begin{eqnarray*}
        L_{n} ( \btheta) &=& \Tr \( \hat{\bV}_H ^{-1}  \frac{1} {2n}\sum _{k=1} ^{n}  \left \{ \vech(IV_k) -  \tau \bh_{h,k}^H (\btheta ) \right \}   \left \{ \vech(IV_k) -  \tau \bh_{h,k}^H (\btheta ) \right \} ^{\top}  \)  \cr
        &&  +  \Tr \( \hat{\bV}_L ^{-1}  \frac{1} {2n}\sum _{k=1} ^{n}  \left \{\br_{h,k}^2 (\bmu_{L}) -  (1-\tau) \bh_{h,k}^L (\btheta ) \right \}   \left \{ \br_{h,k}^2 (\bmu_{L}) - (1- \tau) \bh_{h,k}^L (\btheta ) \right \} ^{\top} \).  
        \end{eqnarray*}
    Thus, we have
        \begin{eqnarray*}
        L_{n} ( \btheta) -L_{0, n}   (\btheta)  &=& \Tr \( \hat{\bV}_H ^{-1}  \frac{1} {2n}\sum _{k=1} ^{n}   D_{0,H,k} (\btheta) \)  +  \Tr \( \hat{\bV}_L ^{-1}  \frac{1} {2n}\sum _{k=1} ^{n}   D_{0,L,k}(\btheta)  \),   
        \end{eqnarray*}
    where $D_{0,H,k}(\btheta)$ and $ D_{0,L,k}(\btheta) $ are martingale differences.
    Since $D_{0,H,k}(\btheta)$ and $ D_{0,L,k}(\btheta) $ are martingale differences, the martingale convergence theorem shows, for any given $\btheta \in \bTheta$, 
        \begin{eqnarray*}
        L _{n}  (\btheta) -L_{0, n}   (\btheta) = O_p(n^{-1/2}). 
        \end{eqnarray*}
    Moreover, similar to the proofs of Lemma 3 in \citet{kim2016unified}, we can show that $L _{n}  (\btheta) -L_{0, n}   (\btheta)$ is stochastic equicontinuous.
    Thus, Theorem 3 in \citet{andrews1992generic} implies the uniformly convergence.
    \endpf

    \begin{thm}\label{Thm-consist}
    Under the assumptions in Theorem \ref{Thm:Theta-Large}, for $  n \geq d$, there is a unique maximizer of $L_{0,n}  (\btheta)$ and $\hat{\btheta}  -\btheta_0 = o_p(1)$.
    \end{thm}
    
    \textbf{Proof Theorem \ref{Thm-consist}.}
    We can easily show that $\btheta^*$ is a maximizer of $L_{0,n} (\btheta)$, if $\bh_{i}^H (\btheta ^{*}) =\bh_{i}^H (\btheta_0)$  and $\bh_{i}^L (\btheta ^{*}) =\bh_{i}^L (\btheta_0)$ for $i \in \{1, \ldots, n \}$. 
    Since $\bX(t)$'s are nondegenerating,  the solution is unique almost surely. 
    Thus, there exists a unique maximizer of $L_{0,n}  (\btheta)$. 
    Finally, Theorem 1 in \citet{xiu2010quasi} provides  the consistency of $\hat{\btheta}$ from Lemma \ref{Lemma:L-consistent} and the identifiability of $\btheta_0$.  
    \endpf

    \textbf{Proof of Theorem \ref{Thm:Theta-Large}.}    
    Taylor expansion and mean value theorem provide that for some $\tilde{\btheta}$ between $\btheta_0$ and $\hat{\btheta} $, we have
    $$\hat{\psi} _{n,m}   (\hat{\btheta} )-\hat{\psi} _{n,m}   (\btheta _0)= -\hat{\psi} _{n,m}  (\btheta _0)=\triangledown \hat{\psi} _{n,m}  (\tilde{\btheta}) (\hat{\btheta}  -\btheta _0).$$
    If $\triangledown \hat{\psi} _{n,m}  (\tilde{\btheta}) -\triangledown \psi _{n}   (\btheta_0) =o_p(1)$, where $-\triangledown \psi _{n}   (\btheta_0)$ is positive definite, then, the convergence rate of $(\hat{\btheta}  -\btheta _0)$ is the same as that of $\hat{\psi} _{n,m}  (\btheta _0)$.
    First, let us investigate the convergence rate of $\hat{\psi} _{n,m}  (\btheta _0)$. 
    A straightforward calculation yields
        \begin{eqnarray*}
        \frac{\partial \hat{l}_{i}  (\btheta_0)}{\partial \theta_k} &=&   - 2 \tau  \left \{ \vech(RV_i) -  \tau\hat{\bh}_{h,i}^H (\btheta_0 ) \right \} ^{\top} \hat{\bV}_H ^{-1}  \frac{\partial \hat{\bh}_{h,i}^H (\btheta_0)}{\partial \theta_k}   \cr
        &&-2    \left \{\hat{\br}_{h,i}^2 (\bmu_{L0}) -  (1-\tau) \hat{\bh}_{h,i}^L (\btheta_0 ) \right \}^{\top} \hat{\bV}_L ^{-1}  \frac{\partial \hat{\br}_{h,i}^2 (\bmu_{L0}) -  (1-\tau) \hat{\bh}_{h,i}^L (\btheta_0 )}{\partial \theta_k}
        \end{eqnarray*}
    and 
        \begin{eqnarray*}
        \frac{\partial l_{i}  (\btheta_0)}{\partial \theta_k} &=&   - 2 \tau  \left \{ \vech(IV_i^f) -  \tau\bh _{h,i}^H (\btheta_0 ) \right \} ^{\top}  \hat{\bV}_H ^{-1} \frac{\partial  \bh _{h,i}^H (\btheta_0)}{\partial \theta_k}   \cr
        &&-2   \left \{\br_{h,i}^2 (\bmu_{L0}) -  (1-\tau)  \bh _{h,i}^L (\btheta_0 ) \right \} ^{\top} \hat{\bV}_L ^{-1}  \frac{\partial \br_{h,i}^2 (\bmu_{L0}) -  (1-\tau)  \bh _{h,i}^L (\btheta_0 )}{\partial \theta_k}.
        \end{eqnarray*}
By \eqref{Thm-large:eq-01} and \eqref{Thm-large:eq-03}, we can show
        \begin{eqnarray*}
        && \| \hat{\bh}_{h,i}^H (\btheta_0 ) - \bh_{h,i}^H (\btheta_0 )\|_{\max} = O_p(m^{-1/4}+\nu_{m,n}+ \pi(p)/p), \cr
        &&  \| \hat{\bh}_{h,i}^L (\btheta_0 ) - \bh_{h,i}^L (\btheta_0 )\|_{\max} = O_p(m^{-1/4}+\nu_{m,n}+ \pi(p)/p), \cr
        && \left \| \frac{\partial\hat{\bh}_{h,i}^H (\btheta_0 )}{\partial{\theta_k} } - \frac{\partial \bh_{h,i}^H (\btheta_0 )} {\partial \theta_k}  \right \|_{\max} = O_p(m^{-1/4}+\nu_{m,n}+ \pi(p)/p),  \cr
        &&\left \| \frac{\partial\hat{\bh}_{h,i}^L (\btheta_0 )}{\partial{\theta_k} } - \frac{\partial \bh_{h,i}^L (\btheta_0 )} {\partial \theta_k}  \right \|_{\max} = O_p(m^{-1/4}+\nu_{m,n}+ \pi(p)/p).
        \end{eqnarray*}
    Thus, we have
        \begin{equation*}
        \left | \frac{\partial \hat{l}_{i}  (\btheta_0)}{\partial \theta_k} -
        \frac{\partial l_{i}  (\btheta_0)}{\partial \theta_k} \right | = O_p (m^{-1/4}+\nu_{m,n}+ \pi(p)/p).
        \end{equation*}
    Now consider $\psi  _{n}  (\btheta _0)$.
    We have
        \begin{eqnarray*}
        \psi  _{n}  (\btheta _0)   &=&   - 2 \frac{1}{n} \sum _{i=1}^n  \Big [\frac{\partial  \tau \bh _{h,i}^H (\btheta_0)}{\partial \btheta}   \hat{\bV}_H ^{-1}  \bQ_{h,i} ^H    + \frac{\partial \br_{h,i}^2 (\bmu_{L0}) -  (1-\tau)  \bh _{h,i}^L (\btheta_0 )}{\partial \btheta }   \hat{\bV}_L ^{-1}  \bQ_{h,i} ^{LL}   \Big ] \cr
        &=& - \frac{ 2 }{n} \Tr \( \hat{\bV}_H ^{-1}  \sum _{i=1}^n    \bQ_{h,i} ^H  \frac{\partial  \tau \bh _{h,i}^H (\btheta_0)}{\partial \btheta} + \hat{\bV}_L ^{-1}   \sum _{i=1}^n  \bQ_{h,i} ^{LL}  \frac{\partial \br_{h,i}^2 (\bmu_{L0}) -  (1-\tau)  \bh _{h,i}^L (\btheta_0 )}{\partial \btheta } \) .
        \end{eqnarray*}
    Since $ \bQ_{h,i} ^H $ and $ \bQ_{h,i} ^{LL}$ are martingale differences,  the martingale convergence theorem shows 
        $$
        \psi  _{n }  (\btheta _0) = O_p(n^{-1/2}).
        $$
    Therefore, we have
        $$
        \hat{\psi}  _{n,m}  (\btheta _0) = O_p(m^{-1/4} + n^{-1/2} +\nu_{m,n}+ \pi(p)/p).
        $$
    Now, it is enough to show that  $\triangledown \hat{\psi} _{n,m}  (\tilde{\btheta}) -\triangledown \psi _{n}   (\btheta_0) =o_p(1)$. This can be showed by the same argument to the proof of Theorem 2 \citep{kim2016unified} using Theorem \ref{Thm-consist}. 
    \endpf

    
    \subsection{Proof of Theorem \ref{Thm-LargePredict}}
                
    \textbf{Proof of Theorem \ref{Thm-LargePredict}.}
    The statements \eqref{Thm-LargePredict:result-1} and \eqref{Thm-LargePredict:result-2} can be showed by Theorem 4.1 \citep{fan2018robust}.

    Consider \eqref{Thm-LargePredict:result-3}. 
    We have
        \begin{eqnarray*}
        \| \tilde{\bGamma}_{n+1} - \E\( \bGamma_{n+1} | \FF_{n} \) \| _{\bGamma^*}
        & \leq&  \| \hat{\bU} \hat{\bH}_n (\hat{\btheta}) \hat{\bU}^{\top}  - \bU   \bH _n ( \btheta_0)  \bU ^{\top}  \| _{\bGamma^*} + \| \hat{\bGamma}  ^{s} - \bGamma  ^s \|_2 \cr
        &\leq& \| \hat{\bU} \hat{\bH}_n (\hat{\btheta}) \hat{\bU}^{\top}  - \bU   \bH _n ( \btheta_0)  \bU ^{\top}  \| _{\bGamma^*} +O_p\(  \pi(p)   b_{m,n}^{1-\delta} \),
        \end{eqnarray*}
    where the last inequality is due to \eqref{Thm-LargePredict:result-2}.
    Now, we need to show 
        \begin{eqnarray*}
        \| \hat{\bU} \hat{\bH}_n (\hat{\btheta} ) \hat{\bU}^{\top}  - \bU   \bH _n ( \btheta_0)  \bU ^{\top}  \| _{\bGamma^*} &=& O_p \Big (  m^{-1/4} + n^{-1/2} + \nu_{m,n} +   \frac{\pi(p)}{ p^{1/2}} \cr
        && \qquad \qquad    +p^{1/2} (m^{-1/2} + n^{-1} + \nu_{m,n}^2)+ \pi(p)  b_{m,n}^{1-\delta} \Big ).
        \end{eqnarray*}
    Using Theorem \ref{Thm:Theta-Large}, we can easily show that 
        \begin{eqnarray*}
        &&\|\hat{\bomega} ^g- \bomega^g_0 \|_2 = O_p \(  m^{-1/4}+n^{-1/2}+\nu_{m,n}+ (\pi(p)/p)^{1/2} \),   \cr
        &&\| \hat{\bR}^g - \bR^g_0 \|_2 = O_p \(  m^{-1/4}+n^{-1/2}+\nu_{m,n}+ (\pi(p)/p)^{1/2} \), \cr
        &&\| \hat{ \bA}^g - \bA^g_0 \|_2 = O_p \(  m^{-1/4}+n^{-1/2}+\nu_{m,n}+ (\pi(p)/p)^{1/2} \), \cr
        &&\| \hat{ \bB}^g - \bB^g_0 \|_2 = O_p \(  m^{-1/4}+n^{-1/2}+\nu_{m,n}+ (\pi(p)/p)^{1/2} \).
        \end{eqnarray*}
    Then, similar to the proofs of Theorem \ref{Thm:Theta-Large}, we can show
        \begin{eqnarray}
        \label{Thm-LargePredict:eq-1}
        \| \hat{\bH} _n (\hat{\btheta}) - \bH_n (\btheta_0) \|_F = O_p \(m^{-1/4}+n^{-1/2}+\nu_{m,n}+ (\pi(p)/p)^{1/2} \).
        \end{eqnarray}
    Assumption \ref{Assumption1}(d)  and \eqref{Thm-LargePredict:eq-1} show
        \begin{eqnarray*}
        \| \hat{\bU} \hat{\bH} _n (\hat{\btheta} ) \hat{\bU}^{\top} -\bU \bH_n (\btheta_0)  \bU^{\top}
        \|_F = O_p \(p \{m^{-1/4}+n^{-1/2}+ \nu_{m,n}+ (\pi(p)/p)^{1/2}\} \).
        \end{eqnarray*}
    Similar to the proofs of Theorem 4.1 \citep{fan2018robust}, we can show
        \begin{eqnarray*}
        \| \hat{\bU} \hat{\bH} _n (\hat{\btheta} ) \hat{\bU}^{\top} -\bU \bH_n (\btheta_0)  \bU^{\top}
        \|_{\bGamma^*} &\leq& C \Big [ p^{-1} \|  \hat{\bU} \hat{\bH} _n (\hat{\btheta} ) \hat{\bU}^{\top} -\bU \bH_n (\btheta_0)  \bU^{\top}\|_F \cr
        && \qquad + p^{-3/2} \|  \hat{\bU} \hat{\bH} _n (\hat{\btheta} ) \hat{\bU}^{\top} -\bU \bH_n (\btheta_0)  \bU^{\top}\|_F^2 \Big ] \cr
        &=&  O_p \Big (  m^{-1/4} + n^{-1/2} + \nu_{m,n} +   \frac{\pi(p)}{ p^{1/2}} \cr
        && \qquad \qquad    +p^{1/2} (m^{-1/2} + n^{-1} + \nu_{m,n}^2)+ \pi(p)  b_{m,n}^{1-\delta} \Big ).
        \end{eqnarray*}
    Thus, we have
        \begin{eqnarray*}
        \| \tilde{\bGamma}_{n+1} - \E\( \bGamma_{n+1} | \FF_{n} \) \| _{\bGamma^*} &=& O_p \Big (  m^{-1/4} + n^{-1/2} + \nu_{m,n} +   \frac{\pi(p)}{ p^{1/2}} \cr
        && \qquad \qquad    +p^{1/2} (m^{-1/2} + n^{-1} + \nu_{m,n}^2)+ \pi(p)  b_{m,n}^{1-\delta} \Big ).
        \end{eqnarray*}
    \endpf
 
 \newpage

\section{Notation table}

\begin{table}[!htp]
\caption{Definitions of notations. We note that the hat notation is used for the estimator.  }
\begin{tabular}{lllll} \hline
 Notation & Meaning     \\ \hline
 $\bX (t)$& true log price process   \\
 $\bff (t)$ & latent factor process    \\
  $\bu (t)$& latent idiosyncratic process  \\
 $\bsigma_t$ &    factor volatility process matrix    \\
 $\bvartheta_t$ &    idiosyncratic volatility process matrix   \\
  $\bU$ &  factor loading matrix    \\
  $\bmu_{0}$ & drift vector  \\
 $\bB_t$,  $\bW_t$, $\tilde{\bB}_t$  &  Brownian motions   \\
$\bSigma_t$ & $\bsigma_t ^{\top} \bsigma_t $   \\
	$\bGamma_{n}$ & daily integrated volatility matrix   \\
  $IV_k^f , \bzeta_n$  & daily factor volatility matrix \\
 $\bGamma_{n}^s$ & daily idiosyncratic volatility matrix  \\
  $\br_t$ & factor return   \\
   $\pi(p)$   & sparsity level  \\
$\tau$  &  length of the market open-to-close period     \\
$\bomega_{H1}, \bomega_{H2}, \bomega_{L1}, \bomega_{L2},  \bnu, \bbeta_H,\bbeta_L, \bgamma_H, \bgamma_L$    & model parameters   \\
$\bh_n^H (\btheta) $  & conditional volatility for open-to-close    \\
$\bh_n^L (\btheta) $  &  conditional volatility for close-to-open     \\
$\bh_n (\btheta) $  &  conditional volatility for the whole-day    \\
 $\bomega_{H}^g,  \bR_H^g  ,    \bA_H^g,   \bB_H^g$ &  GARCH parameters for open-to-close   \\
  $\bomega_{L}^g,  \bR_L^g  ,    \bA_L^g,   \bB_L^g$ &  GARCH parameters for close-to-open    \\
    $\bomega^g,  \bR ^g  ,    \bA ^g,   \bB ^g$ &  GARCH parameters for whole-day  \\
  $  \bQ_{n}^H$ &  noise for open-to-close volatility    \\  
    $  \bQ_{n}^L$ &   noise for close-to-open volatility   \\  
        $  \bQ_{n}^{LL}$ &  noise for squared close-to-open return   \\  
    $ Y _{i}(t_{k, \ell})$ &  noisy observation   \\  
    $ \epsilon _{i}( t_{k, \ell})$ &  micro-structural noise  \\  
  $d$ &   the cardinality of $\btheta$  \\  
    $p$ &   the number of assets   \\  
        $r$ &   the rank  of factor volatility  \\  
 $  RV_k$ &   estimator of latent daily factor volatility   \\  
  $\bV_H$ &  covariance matrix for the $\bQ_n^{H}$    \\  
 $\bV_L$&   covariance matrix for the $\bQ_n^{LL}$  \\  
  $\lambda_{n,i}$ &   the $i$th largest eigenvalue of $\bU \bPsi_n \bU ^{\top}$   \\  
   $ \bGamma ^{*}$ &   $ \E\( \bGamma_{n+1} | \FF_{n} \)$    \\  \hline
\end{tabular}
\end{table}

\bibliography{myReferences}

\end{spacing}